\documentclass[11pt,a4paper]{article}
\usepackage{jheppub-liu}
\usepackage{bbm}
\usepackage{braket}
\usepackage{multirow}
\usepackage{empheq}

\begin{document}

\makeatletter
\gdef\@fpheader{\sc Prepared for submission to JHEP 
}
\makeatother

\title{Scattering Equations, Twistor-string Formulas \\and Double-soft Limits in Four Dimensions}
\author[a]{Song He}
\author[b,c]{Zhengwen Liu}
\author[d,e]{and Jun-Bao Wu}

\affiliation[a]{Key Laboratory of Theoretical Physics,\\
Institute of Theoretical Physics, Chinese Academy of Sciences,
Beijing 100190, P. R. China}

\affiliation[b]{Center for Cosmology, Particle Physics and Phenomenology (CP3), \\
Universit\'{e} catholique de Louvain, B1348 Louvain-la-Neuve, Belgium}

\affiliation[c]{Department of Physics, Renmin University of China, Beijing 100872, P. R. China}

\affiliation[d]{Institute of High Energy Physics, and Theoretical Physics Center for Science Facilities,\\
Chinese Academy of Sciences, 19B Yuquan Road, Beijing 100049, P. R. China}

\affiliation[e]{Center for High Energy Physics, Peking University, No. 5 Yiheyuan Road, Beijing 100871, P. R. China
}

\emailAdd{songhe@itp.ac.cn}
\emailAdd{zhengwen@ruc.edu.cn}
\emailAdd{wujb@ihep.ac.cn}

\abstract
{We study scattering equations and formulas for tree amplitudes of various theories in four dimensions, in terms of spinor helicity variables and on-shell superspace for supersymmetric theories. As originally obtained in Witten's twistor string theory and other twistor-string models, the equations can take either polynomial or rational forms, and we clarify the simple relation between them. We present new, four-dimensional formulas for all tree amplitudes in the non-linear sigma model, a special Galileon theory and the maximally supersymmetric completion of the Dirac-Born-Infeld theory. Furthermore, we apply the formulas to study various double-soft theorems in these theories, including the emissions of a pair of soft photons, fermions and scalars for super-amplitudes in super-DBI theory.
}
\arxivnumber{1604.02834}
\maketitle

\section{Introduction}
In a series of works, a new formulation has been developed, which expresses tree-level S-matrix of massless particles in arbitrary dimensions, as an integral over the moduli space of Riemann spheres. This so-called Cachazo-He-Yuan (CHY) representation has been proposed originally for amplitudes in gravity, Yang-Mills, and bi-adjoint scalar theories~\cite{Cachazo-He-Yuan-1307.2199, Cachazo-He-Yuan-1309.0885}, and extended to a large variety of theories in~\cite{Cachazo-He-Yuan-1409.8256, Cachazo-He-Yuan-1412.3479}. For example, a remarkably simple formula gives all multi-trace gluon-graviton amplitudes in the Einstein-Yang-Mills theory; other theories
with compact CHY formulas include the $U(N)$ non-linear sigma model (NLSM), Dirac-Born-Infeld (DBI) theory and a special Galileon theory\footnote{Galileon theories are effective field theories in the decoupling limit of massive gravity~\cite{Hinterbichler:2011tt} and DGP model \cite{Dvali:2000hr}. Amplitudes in these theories have been studied in e.g.~\cite{Kampf:2014rka}.
} (sGal)~\cite{Cachazo-He-Yuan-1412.3479,Hinterbichler-Joyce-1501.07600}.

Different theories correspond to different integrands of the integral formula, but the universal part in the construction is given by the delta-function constraints, known as scattering equations~\cite{Cachazo-He-Yuan-1306.2962, Cachazo-He-Yuan-1306.6575, Cachazo-He-Yuan-1307.2199}
\begin{align}
  \sum_{b\neq a} \frac{k_a\cdot k_b}{\sigma_{a}-\sigma_{b}} \,=\, 0,
  \quad \text{for}~a=1, 2, \ldots, n,
  \label{scatt}
\end{align}
where $\sigma_a$ denotes the position of the $a^{\rm th}$ puncture. Only $n-3$ equations out of the $n$ equations are independent because of the $\operatorname{SL}(2,{\mathbb C})$ symmetry, and the system has $(n-3)!$ solutions in general. These equations have made an appearance in previous literature in different contexts~\cite{Fairlie:1972, *Roberts:1972, Fairlie-0805.2263, Gross:1987ar, Witten-0403199, Makeenko-Olesen-1111.5606, Cachazo-1206.5970}. Elegant worldsheet models have been proposed~\cite{Mason-Skinner-1311.2564, Berkovits:2013xba} for the original CHY formulas, and more recently generalized to these new theories~\cite{Ohmori-1504.02675, Casali:2015vta}.

In four dimensions, further simplifications occur since any null vector can be written as a bi-spinor, $k^{\alpha \dot{\alpha}}=\lambda^\alpha \tilde\lambda^{\dot\alpha}$. As first pointed out in~\cite{Cachazo-He-Yuan-1306.2962}, when reduced to 4d the scattering equations become delta-function constraints of Roiban-Spradlin-Volovich (RSV) formula for ${\cal N}\!=\!4$ super-Yang-Mills (SYM) tree amplitudes~\cite{Roiban-Spradlin-Volovich-0403190}, originally derived from Witten's twistor string theory~\cite{Witten-0312171}.
These equations also appeared in two different formulas for ${\cal N}\!=\!8$ supergravity (SUGRA) tree amplitudes, proposed in~\cite{Cachazo-Geyer-1206.6511, Cachazo-Skinner-1207.0741} and later derived from a new twistor string theory~\cite{Cachazo-Mason-Skinner-1207.4712, Skinner-1301.0868}.
We will refer to them as four-dimensional {\it polynomial scattering equations} since they take polynomial form with degree $d=1,\ldots, n{-}3$:
\begin{align}\label{4dp}
  \sum_{a=1}^n t_a \sigma_a^m \tilde\lambda^{\dot{\alpha}}_a \,=\, 0
  \quad\text{for}~m=0,1,\ldots,d,
  \qquad
  \lambda_a^\alpha - t_a \sum_{m=0}^d \rho_m^\alpha \sigma_a^m \,=\, 0
  \quad\text{for}~a=1,\ldots,n,
\end{align}
where the variables are $\sigma_a, t_a$ for $a=1,2,\ldots,n$ and $\rho^{\alpha}_m$ for $m=0,\ldots, d$, and the scattering equations decompose into $n-3$ sectors labeled by $d$. It is well known that for both Yang-Mills and gravity case, exactly the sector-$d$ equations are needed for amplitudes in the helicity sector $k=d+1$ (e.g.~those with $k$ negative-helicity gluons or gravitons). Importantly, as derived in~\cite{Cachazo-He-Yuan-1306.2962}, the number of solutions is the Eulerian number~$E_{n{-}3,d{-}1}$~for the 4d scattering equations in sector $d$, which add up to the total number of solutions $(n{-}3)!=\sum_{d=1}^{n{-}2} E_{n{-}3,d{-}1}$.

In~\cite{Geyer-Lipstein-Mason-1404.6219}, based on the four-dimensional ambitwistor string theory, similar formulas have been obtained for ${\cal N}\!=\!4$ SYM and ${\cal N}\!=\!8$ SUGRA amplitudes.
The delta-function constraints in these formulas are labeled by $k=2,\ldots, n{-}2$ and take rational form (hence will be referred to as four-dimensional {\it rational scattering equations}). It is convenient to divide $n$ particles into two sets, one of $k$ particles and the other of $n{-}k$, e.g.~$\{1,2,\ldots, k\}$ and $\{k{+}1,k{+}2,\ldots,n\}$, and the rational form of the equations reads
\begin{align}\label{4dr}
  \tilde\lambda_I^{\dot{\alpha}}  - \sum_{i=k+1}^n {\tilde\lambda^{\dot{\alpha}}_i \over (I\,i)} \,=\, 0
  ~~\text{for}~I=1,\ldots,k,
  \quad
  \lambda_i^\alpha - \sum_{I=1}^k {\lambda_I^\alpha \over (i\,I)} \,=\, 0
  ~~\text{for}~i=k+1,\ldots,n,
\end{align}
where we have defined the two-bracket $\displaystyle (a\,b) := (\sigma_a-\sigma_b)/(t_at_b)$ by writing $\sigma$'s as $\sigma_a^\alpha = {1\over t_a}(1, \sigma_a)$.  As discussed in~\cite{Geyer-Lipstein-Mason-1404.6219}, the formulas based on these rational equations have a relatively simpler form, but with the bose/fermi symmetry not as manifest as those with polynomial equations.

In the first part of the paper, we study four-dimensional tree amplitudes in various theories with these 4d scattering equations. As we will review shortly, both forms of the 4d equations can be derived from reducing the scattering equations \eqref{scatt} to four dimensions, so they are of course equivalent to each other. However, since the twistor string theories behind these two types of formulas are very different, it is not obvious at all how to directly connect these them. In section \ref{sec2.1}, We will show that the two forms of the 4d scattering equations, \eqref{4dp} and \eqref{4dr}, are simply related to each other by a GL$(k)$ transformation. It turns out that \eqref{4dr} is basically the GL$(k)$-fixed version of \eqref{4dp} with e.g.~$\{1,2,\ldots, k\}$ chosen as the labels for fixing the GL$(k)$ redundancies. When imposing these equations in delta functions of these twistor-string-inspired formulas (or 4d CHY formula in short), we work out the action of GL$(k)$ transformation on the formula, then it becomes obvious how to translate between the integral measure and integrand of the formula with polynomial or rational form equations.

We proceed in section \ref{sec-4d-formulas} to write new 4d formulas for tree amplitudes in NLSM, DBI and sGal. As pointed out in~\cite{Cachazo-He-Yuan-1412.3479}, a significant simplification is that only the middle sector, $k=d{+}1=n/2$, is needed for amplitudes in these theories. While NLSM and sGal only have scalars, we find that the formula for DBI amplitudes begs to be put in the supersymmetric form when reduced to four dimensions. This is parallel to the cases of Yang-Mills and gravity formulas in 4d, which take the nicest form as we include the supermultiplet and write them in a manifestly supersymmetric manner~\cite{Roiban-Spradlin-Volovich-0403190, Cachazo-Geyer-1206.6511, Cachazo-Skinner-1207.0741}. In the end, the formula naturally leads us to find the ${\cal N}=4$ supersymmetric completion of the usual DBI theory. Together with the formula for bi-adjoint $\phi^3$ theory~\cite{Cachazo-Zhang-1601.06305} and that for Einstein-Yang-Mills amplitudes in 4d (and its supersymmetric extensions)~\cite{Adamo-Casali-Roehrig-Skinner-1507.02207}, all CHY formulas discovered so far have been written as twistor-string-like form in 4d, with supersymmetric extensions when possible.

The upshot is a nice formula for all super-amplitudes in the maximally (${\cal N}\!=\!4$) supersymmetric DBI theory, for which we will refer as super-DBI or SDBI for short~\cite{Tseytlin-9908105}. This is a theory with half of the supersymmetries linearly realized and half non-linearly realized, and the Lagrangian of the theory has only been written down very recently~\cite{Bergshoeff-Coomans-Kallosh-Shahbazi-Proeyen-1303.5662}. The fermionic sector is known to coincide with the Volkov-Akulov theory~\cite{Volkov-Akulov-1972} with fermions now carrying fundamental SU(4) indices. We find it intriguing that a very compact formula contains all tree-level amplitudes of the theory with such a complicated Lagrangian. In fact, as already expected from \cite{Cachazo-He-Yuan-1412.3479}, only one new ingredient is needed to get our 4d formula for amplitudes in SDBI, NLSM and sGal.

In the second part we apply our results to a very interesting problem:~emission of soft particles for amplitudes in these theories. There has been renewed interests in exploring connections between symmetries and universal soft behavior of amplitudes~\cite{Cachazo-Strominger-1404.4091}. Here we will focus on the emission of soft Goldstone particles of spontaneously broken symmetry. The famous Adler's zero means that the emission of a single soft Goldstone boson gives vanishing amplitude~\cite{Adler-1965, Weinberg-1966-Pino-Scattering}, and double-soft emission probes the coset algebra structure of the vacuum (c.f.~\cite{Arkani-Hamed-Cachazo-Kaplan-0808.1446} for double-soft-scalar emission in ${\cal N}\!=\!8$ SUGRA). More recently, new double-soft-emission theorems have been proposed for certain effective field theories with spontaneously broken symmetries, such as NLSM, DBI and sGal~\cite{Cachazo-He-Yuan-1503.04816}. Double-soft-fermion emission has been studied in various theories including Volkov-Akulov theory and SUGRA~\cite{Chen-Huang-Wen-1412.1809}.

In section \ref{sec3} we use our four-dimensional formulas to obtain double soft emissions of the theories under consideration, including scalar-emissions in NLSM and sGal, and emissions of scalars, fermions and other particles in SDBI, SYM and SUGRA. The 4d formulas allow us to derive all these universal double-soft theorems; in particular, when the flavors of the two soft particles do not form a SU(${\cal N}$) singlet, the leading order vanishes and we obtain sub-leading soft theorems probing coset structure of broken symmetries. Note that for these effective field theories, it is not clear how to apply standard techniques such as BCFW recursions~\cite{Britto-Cachazo-Feng-0412308, Britto-Cachazo-Feng-Witten-0501052}, thus it is important that our formula provides very strong evidence in favor of these theorems. The double-soft theorems in super-DBI theory are particularly interesting since they will provide clues for the mysterious non-linearly realized (super)symmetries of the theory.

\section{Scattering equations and formulas in four dimensions}

\subsection{Relations between different forms of 4d scattering equations}\label{sec2.1}

We start with a lightening review of how \eqref{4dp} and \eqref{4dr} follow from \eqref{scatt} when reduced to four dimensions. For \eqref{4dp} it is more convenient to go to the manifestly parity-invariant form~\cite{Witten-0403199, Cachazo-1301.3970}, which can be obtained by introducing $\tilde\rho$'s and rewriting equations for $\tilde\lambda$'s similar to those for $\lambda$'s~\cite{Cachazo-1301.3970}:
\begin{align}\label{4dp2}
\lambda_a - t_a \sum_{m=0}^d \rho_m \sigma_a^m \,=\, 0,\quad \tilde\lambda_a - \tilde t_a \sum_{m=0}^{\tilde d} \tilde\rho_m \sigma_a^m\,=\, 0, \quad t_a \tilde t_a=\prod_{b\neq a} \frac 1 {\sigma_a-\sigma_b}\,,
\end{align}
where $\tilde d=n{-}2{-}d$. These equations are completely equivalent to \eqref{4dp}, and it was first shown in~\cite{Cachazo-He-Yuan-1306.2962} (see also~\cite{Cachazo-Zhang-1601.06305}) that their solutions of all the sectors $d=1,\ldots, n{-}3$ are in one-to-one correspond to those of scattering equations when reduced to four dimensions\,\footnote{By this we mean one-to-one correspondence between solutions of \eqref{scatt} for $\sigma$'s and solutions for $\sigma$'s of \eqref{4dp} (or \eqref{4dr}) with all sectors. Given any such $\sigma$-solution, the solutions for $t$'s in \eqref{4dr} (and $\rho$'s, $t$'s in \eqref{4dp}) are determined uniquely by the equations.}.

The scattering equations, \eqref{scatt}, were originally derived as the null condition $p^2(z)=0$ for a vector-valued polynomial map from Riemann sphere to momentum space:~\cite{Cachazo-He-Yuan-1306.6575}:
\begin{align}\label{poly}
  p^\mu(z) \,:=\, \sum_{a=1}^n\, k_a^\mu~\prod_{b\neq a}\,(z-\sigma_b)\,,
\end{align}
which is a degree-$(n-2)$ polynomial. In four dimensions, $p^2(z)=0$ is equivalent to the existence of polynomials $\lambda(z):=\sum_{m=0}^d \rho_m z^m$ and $\tilde\lambda(z):=\sum_{m=0}^{\tilde d} \tilde\rho_m z^m$, such that $p^{\alpha\dot\alpha}(z)=\lambda^\alpha(z) \tilde\lambda^{\dot\alpha}(z)$. This is the origin of sectors in 4d:~the degrees of $\lambda(z)$ and $\tilde\lambda(z)$, $d$ and $\tilde d$ respectively, must satisfy $d + \tilde d= n-2$, thus the solutions of scattering equations must split into exactly $n{-}3$ sectors, $d=1,2,\ldots, n{-}3$. By using \eqref{4dp2} and $k^{\alpha\dot\alpha}_a=\lambda^\alpha_a \tilde\lambda^{\dot \alpha}_a$, we can verify that \eqref{poly} gives $p^{\alpha\dot\alpha}(z)=\lambda^{\alpha} (z) \tilde\lambda^{\dot\alpha} (z)$.

Here we show that the same is true for \eqref{4dr}, and we first define a rational map equivalent to $p^\mu (z)$
\begin{align}
w^\mu (z) \,:=\, \sum_{a=1}^n \frac{k_a^\mu}{z-\sigma_a}
\,=\, {p^\mu(z) \over \prod_{b=1}^n (z-\sigma_b)}\,.
\end{align}
The proof is actually one line:~by plugging \eqref{4dr} into $w^{\alpha \dot{\alpha}}$ and recall that $(I\,i)=(\sigma_I-\sigma_i)/(t_I t_i)$:
\begin{align}
  w^{\alpha \dot{\alpha}}
  = \sum_{I=1}^k \frac{\lambda^\alpha_I \tilde\lambda^{\dot{\alpha}}_I}{z-\sigma_I}
  +\sum_{i=k{+}1}^n \frac{\lambda^\alpha_i\tilde\lambda^{\dot{\alpha}}_i}{z-\sigma_i}
  = \sum_{I,i} \frac{t_It_i\,\lambda^\alpha_I\tilde\lambda^{\dot{\alpha}}_i}{\sigma_I-\sigma_i}
  \left(\frac 1{z-\sigma_I}-\frac 1 {z-\sigma_i}\right)
  = \left(\sum_{I=1}^k \frac{ t_I \lambda^\alpha_I}{z-\sigma_I}\right)
  \left(\sum_{i=k{+}1}^n {t_i\tilde\lambda_i^{\dot\alpha}\over z-\sigma_i}\right),
  \nonumber
\end{align}
which immediately gives $w^2(z)=0$. Thus any solution of \eqref{4dr} is a solution of $w^2(z)=0$, or equivalently the scattering equations \eqref{scatt}. Since the total number of solutions from all sectors of \eqref{4dr} (or {\eqref{4dp2}) is $(n{-}3)!$, we see that any solution of \eqref{scatt} also corresponds to a solution of \eqref{4dr} (or \eqref{4dp2}).

Now we turn to the transformation between the two forms of 4d scattering equations. It was first pointed out in~\cite{Arkani-Hamed-Bourjaily-Cachazo-Trnka-0912.4912} that \eqref{4dp} can be viewed as the constraints on $\sigma$'s through those on the so-called Veronese form of the Grassmannian ($=k\times n$ matrix up to $\operatorname{GL}(k)$ transformation). From \eqref{4dp}, we see that the form of the matrix (the ``C-matrix") reads~\cite{Arkani-Hamed-Bourjaily-Cachazo-Trnka-0912.4912}
\begin{align}
  C_{m{+}1,a} = t_a \sigma_a^m\,,\quad \text{for}~~
   m=0,\ldots,d, \quad a=1,\ldots, n.
\end{align}
By writing $\lambda,~\tilde\lambda$ both as $n\times 2$ matrices, and $\rho$ as $2\times k$ matrix, \eqref{4dp} become $C\cdot \tilde\lambda=\rho \cdot C-\lambda^{\rm T}=0$ (here the dot ``$\cdot$'' and letter ``${\rm T}$'' denote matrix multiplication and transportation respectively). Geometrically speaking, this means that the $C$-plane is orthogonal to $\tilde\lambda$-plane, and it contains the $\lambda$-plane. For our purpose it is actually more convenient to rewrite the latter constraints as the statement that the orthogonal complement of $C$,  $C^{\perp}$ (which is a $(n{-}k)$-plane or a $(n{-}k) \times n$ matrix), is orthogonal to the $\lambda$-plane. Thus \eqref{4dp} becomes
\begin{align}
C\cdot \tilde\lambda=0\,, \qquad C^{\perp} \cdot \lambda=0\,.
\end{align}

To go from this form to \eqref{4dr} simply requires a $\operatorname{GL}(k)$ transformation $c=L\cdot C$ to bring a $k\times k$ sub-matrix to be the identity.
In our choice, this identity-matrix part is the sub-matrix $c_{I\,J}=\delta_{I\,J}$, and now we need to see what the remaining part, denoted as $c_{I\,i}$ for $i=k+1,\ldots,n$ look like. Note that we have denoted the row labels as $I=1,\ldots, k$. It is straightforward to work out the remaining part, which has been previously spelled out as the link-representation form~\cite{Arkani-Hamed-Bourjaily-Cachazo-Trnka-0903.2110, He-1207.4064}:
\begin{align}\label{link}
  c_{I\,i} \,=\, \frac{t_i \prod_{J\neq I} \sigma_{i J}}{t_I \prod_{K\neq I} \sigma_{I K}}.
\end{align}
Note that after the fixing it is trivial to write $c^{\perp}$ (see below). By performing the transformation
\begin{align}\label{transf1}
  \tilde t_i \,=\, t_i \beta_i,\quad \tilde t_I \,=\, {1\over t_I \beta_I},
  \quad \text{with}\quad \beta_i \,=\, \prod_J \sigma_{iJ},
  \quad \beta_I \,=\, \prod_{K\ne I}\sigma_{IK},
\end{align}
we can absorb an overall factor in \eqref{link}, and the link variables become $c_{I\,i}=\tilde t_I \tilde t_i/\sigma_{I\,i}$.
Let us spell out the constraints $c\cdot \tilde\lambda=c^{\perp}\cdot \lambda=0$ in this gauge-fixed form:
\begin{align}
  \left(\mathbbm{1}_{k\times k}\,|\, c_{k\times (n{-}k)}\right)\cdot \tilde\lambda
  \,=\,0_{k\times 2}\,,\qquad  \left(\left(- c^{\rm T}\right)_{(n{-}k) \times k}\,\big|\, \mathbbm{1}_{(n{-}k) \times (n{-}k)}\right)\cdot\lambda
  \,=\, 0_{(n{-}k)\times 2} \,,
\end{align}
which are exactly \eqref{4dr} (where the $t$'s have been renamed as $\tilde t$'s)! Thus the rational scattering equations, derived from 4d ambitwistor strings in~\cite{Geyer-Lipstein-Mason-1404.6219}, is nothing but the gauge-fixed, or link-representation form of the polynomial equations.

To obtain formulas for tree amplitudes, we need to impose either form of the equations by writing down integral measure localized by delta functions, and when integrating over $d^{2n} \sigma$ there is a $ {\rm GL}(2,{\mathbb C})$ redundancy to be fixed. For theories with ${\cal N}$ supersymmetries, it is very natural to also include fermionic delta functions involving Grassmann odd variables which label the supermultiplet. We use the superspace $(\lambda,\{\tilde\lambda | \eta\})$ with Grassmann variables $\eta^{A}$ where $A=1,\ldots, {\cal N}$, and in this superspace, for example, the on-shell superfields \cite{Nair-1988} for ${\cal N}\!=\!4$ SYM and ${\cal N}\!=\!8$ SUGRA read:
\begin{align}
  \Phi^\text{(SYM)}(\eta)
  \,&=\, g^+ + \eta^A \Gamma_A + {1\over 2!} \eta^{A}\eta^{B}\phi_{AB}
  + {1\over 3!} \eta^{A}\eta^{B}\eta^{C} \epsilon_{ABCD} \bar\Gamma^{D}
  + \eta^{1}\eta^{2}\eta^{3}\eta^{4}\, g^-,
  \label{N=4SYM-supermultiplet}
  \\
  \Phi^\text{(SG)}(\eta)
  \,&=\, h^{+} + \eta^A \lambda_A
  + {1\over 2!} \eta^A \eta^B v_{AB}
  + {1\over 3!} \eta^A \eta^B \eta^C \chi_{ABC}
  + {1\over 4!} \eta^A\eta^B\eta^C\eta^D \phi_{ABCD}
  + \cdots.
  \label{N=8SG-supermultiplet}
\end{align}
Given the superspace, supersymmetry dictates that we include fermonic delta functions for $\eta$'s in the same form as those for $\tilde\lambda$'s. Now we can write down the formulas and see how the measures and integrands of these two forms transform between each other. Let us start with the rational form
\begin{align}\label{}
  {\cal M}_{n, k}&=\int{\prod_{a=1}^n d\sigma_a d\tilde t_a/{\tilde t_a}^3 \over \operatorname{vol} {\rm GL}(2,{\mathbb C})}
  \prod_{I=1}^k \delta^{2|{\cal N}} \left( \{\tilde\lambda_I | \eta_I\} - \sum_{i=k+1}^n {\tilde t_I \tilde t_i\{\tilde\lambda_i|\eta_i\}
  \over \sigma_{Ii}} \right)
  \prod_{i=k+1}^n \delta^{2} \left( \lambda_i
  - \sum_{I=1}^k {\tilde t_i \tilde t_I\lambda_I \over \sigma_{iI}} \right)
  {\cal I}^{\rm rat}_{n,k}  \nonumber\\
  \,&:=\,\int d\mu^{({\cal N})}_{n,k}\,
  {\cal I}^{\rm rat}_{n,k}(\tilde t_i, \tilde t_I)\,,
\end{align}
where ${\cal M}_{n,k}$ is the $n$-point, $k$-sector amplitude in the theory under consideration;\footnote{For NLSM, SDBI and sGal only the sector $k=n/2$ is non-vanishing, and a sum over all sectors is needed for bi-adjoint $\phi^3$ theory; see subsection~\ref{sec-4d-formulas} for details.} on the second line we defined the measure $d\mu^{(\cal N)}_{n,k}$ for rational-form equations with ${\cal N}$ supersymmetries, and (as we will see why shortly) we indicated the explicit dependence of the rational-form integrand ${\cal I}^{\rm rat}$ on $\tilde t_i, \tilde t_I$.

Performing the transformation in~\eqref{transf1} and keeping track of the Jacobians, we get
\begin{align}\label{transformed}
  {\cal M}_{n, k}
  \,&=\,\int {\prod_{a=1}^n d\sigma_ad {t}_a/{t}_a^3 \over \operatorname{vol} {\rm GL}(2,{\mathbb C})}
  \left(\prod_{i=k+1}^n {\beta_i}^{-2}\right)
  \left(\prod_{I=1}^k \beta_{I}^2\,t_I^{4}\right)
  \nonumber\\
  &\qquad\times \prod_{I=1}^k \delta^{2|{\cal N}}
  \left( \{\tilde\lambda_I | \eta_I\} - \sum_{i=k+1}^n {{\beta_i} t_i\{\tilde\lambda_i | \eta_i\}\over \beta_{I} {t}_I\sigma_{Ii}} \right)
  \prod_{i=k+1}^n \delta^{2} \left( \lambda_i
  - \sum_{I=1}^k {{\beta_i} {t}_i\lambda_I \over \beta_{I} {t}_I\sigma_{iI}}\right)
  {\cal I}^{\rm rat}_{n,k} \big({\beta_i} {t}_i, \tfrac{1}{\beta_{I} {t}_I}\big)
  \nonumber\\
  \,&=\,\int d\Omega^{({\cal N})}_{n, k}\,
  (V_k)^{4-{\cal N}}\,
  \left(\prod_{i=k+1}^n {\beta_i}^{-2}\right)
  \left(\prod_{I=1}^k \beta_{I}^2\, {t}_I^{4}\right)\,
  {\cal I}^{\rm rat}_{n,k} \big({\beta_i}{t}_i, \tfrac{1}{\beta_{I} {t}_I}\big).
\end{align}
where the ${\rm GL}(k)$ transformation is performed and we have defined the Jacobian
\begin{align}\label{}
  V_k \,=\, \prod_{I=1}^k {t}_I \prod_{J \neq K} \sigma_{J K},
\end{align}
as well as the polynomial-form measure with ${\cal N}$ supersymmetries ($d^{2n} \sigma = \prod_{a=1}^n d\sigma_a d t_a/ t_a^3$ here)
\begin{align}\label{}
  d\Omega^{({\cal N})}_{n,k}
  \,:=\, {d^{2n}\sigma \over \operatorname{vol} {\rm GL}(2,{\mathbb C})}
  \prod_{m=0}^d \delta^{2|{\cal N}}
  \left( \sum_{a=1}^n t_a\sigma_a^m \{\tilde\lambda_a | \eta_a\}\right)
  \int d^{2k}\rho \prod_{a=1}^n
  \delta^{2} \left( t_a\sum_{m=0}^d \rho_m \sigma_a^m - \lambda_a \right).
\end{align}
From \eqref{transformed}, we find that the integrand with polynomial form of 4d scattering equations is related to the one with rational-form equations in a simple way:
\begin{empheq}[box=\fbox]{align}
  \label{integrands}
  {\cal I}^{\rm pol}_{n,k} (t_a) \,=\, (V_k)^{4-{\cal N}}
  \left(\prod_{i=k+1}^n {\beta_i}^{-2}\right)
  \left(\prod_{I=1}^k \beta_{I}^2\,t_I^{4}\right)\,
  {\cal I}^{\rm rat}_{n,k} \big({\beta_i}t_i, \tfrac{1}{\beta_{I}t_I}\big).
\end{empheq}

\subsection{Formulas for tree amplitudes with 4d scattering equations}\label{sec-4d-formulas}
Now we are ready to write down four-dimensional twistor-string-inspired formulas for tree amplitudes. Note that the formulas contain overall (super-)momentum-conserving delta functions (for supersymmetric theories):~${\cal M}=\delta^4 (P) \delta^{0|2\cal N} (Q) M$ with
$P^{\alpha\dot\alpha}:=\sum_{a=1}^n \lambda^\alpha_a \tilde\lambda^{\dot \alpha}_a$,~$Q^{\dot\alpha, A}:=\sum_{a=1}^n \tilde\lambda_a^{\dot\alpha} \eta_a^A$.

We first recall the twistor-string and ambi-twistor string formulas for  $n$-point N$^{k-2}$MHV, color-ordered tree amplitude in ${\cal N}\!=\!4$ super-Yang-Mills theory (SYM)~\cite{Witten-0312171, Roiban-Spradlin-Volovich-0403190, Geyer-Lipstein-Mason-1404.6219}:
\begin{align}\label{SYM}
  {\cal M}^{\rm SYM}_{n,k}(1,2,\ldots,n)
  \,=\, \int d\Omega^{({\cal N} = 4)}_{n,k}\, {1 \over (12)(23)\cdots(n1)}
  \,=\, \int d\mu^{({\cal N} = 4)}_{n,k}\, {1 \over (12)(23)\cdots(n1)},
\end{align}
From \eqref{integrands}, one can see that the formulas with two forms of the 4d scattering equations have identical integrands. This integrand is just the so-called Parke-Taylor factor.

For ${\cal N} \!=\! 8$ supergravity (SUGRA) amplitudes, the formula with rational equations reads~\cite{Geyer-Lipstein-Mason-1404.6219}:
\begin{align}\label{SUGRA}
  {\cal M}^{\rm SUGRA}_{n,k}
  \,=\, \int d\mu^{({\cal N} = 8)}_{n,k}\,
  {\det}'{\mathbb H}_k\, {\det}'\overline{\mathbb H}_{n{-}k}\,,
\end{align}
where ${\det}'$ denotes the minor with any one column and one row deleted (since the rows and columns add up to zero), and ${\mathbb H}$ and $\overline{\mathbb H}$ are $k\times k$ and $(n-k)\times (n-k)$ matrices of the form:
\begin{align}\label{}
  & {\mathbb H}_{ab} \,=\, {\braket{a\,b} \over (a\,b)}~~\text{for}~a\ne b,
  \quad {\mathbb H}_{aa} \,=\, -\sum_{b=1,b\ne a}^k {\mathbb H}_{ab},
  \quad a,b \in \{1,\ldots, k\};
  \\
  & \overline{\mathbb H}_{ab} \,=\, {[a\,b] \over (a\,b)}~~\text{for}~a\ne b,
  \quad \overline{\mathbb H}_{aa} \,=\, -\sum_{b=k+1,b\ne a}^n \overline{\mathbb H}_{ab},
  \quad a,b \in \{k+1,\ldots, n\}.
\end{align}
Note that the integrand ${\det}'{\mathbb H}_k\, {\det}'\overline{\mathbb H}_{n{-}k}$ is not permutation invariant, but when we rewrite the formula with the polynomial form of the equations, the integrand obtained from (\ref{integrands}) becomes those in~\cite{Cachazo-Geyer-1206.6511, Cachazo-Skinner-1207.0741}, which are permutation invariant.
Henceforth for simplicity we will only write formula with rational form equations explicitly, the formula using polynomial form can be get from \eqref{integrands}.

Very recently the formula for double-partial amplitudes in the bi-adjoint $\phi^3$ theory, ${\cal M}^{\phi^3}_n[\alpha|\beta]$ has been obtained in~\cite{Cachazo-Zhang-1601.06305}. By \eqref{integrands} we translate it into a formula with rational equations:
\begin{align}\label{phi3}
  {\cal M}^{\phi^3}_n [\alpha|\beta]
  \,=\, \sum_{k=2}^{n-2} \int~\frac{d\mu^{(0)}_{n,k}}{{\det}'{\mathbb H}_{k}\,{\det}'\overline{\mathbb H}_{n{-}k}}
~{1\over \alpha[(12)\cdots (n1)]}~{1\over \beta[ (12)\cdots (n1)]},
\end{align}
where we have Parke-Taylor factors with orderings $\alpha,\beta$ and the determinants appeared in \eqref{SUGRA}. It is interesting to see that the formula is more complicated than SYM or SUGRA, especially in that one has to sum over all sectors.
Each $k$ sector gives contributions (the ``scalar blocks"~\cite{Cachazo-Zhang-1601.06305}) with unphysical poles which only cancel each other in the sum over sectors.

The formula for gravity can be derived from ``double-copy" of Yang-Mills, divided by $\phi^3$, which we denote as~``$\text{GR}=\text{YM}\otimes\text{YM}$". This can be viewed as the Kawai-Lewellen-Tye (KLT) relations~\cite{Kawai-Lewellen-Tye-1985} between the amplitudes, or equivalently \cite{Cachazo-He-Yuan-1309.0885, Cachazo-He-Yuan-1412.3479} the relation between CHY integrands of these theories. For example, $\text{GR}=\text{YM}\otimes\text{YM}$ means that, by taking two copies of the CHY integrand for YM, and divided by that of bi-adjoint $\phi^3$ theory, we obtain the CHY integrand for gravity. From the observation of~\cite{Cachazo-Zhang-1601.06305}, a nice feature of the 4d formulas is that this  double-copy procedure works for each $k$-sector individually:~one can easily derive~\eqref{SUGRA} from \eqref{SYM} and \eqref{phi3} for each $k$~\cite{Cachazo-Zhang-1601.06305}.

Now we proceed to formulas for the effective field theories, including super-DBI, NLSM and sGal.
We first consider ${\cal N}\!=\!4$ super-DBI theory, which has an on-shell superfield
\begin{align}\label{N=4-on-shell-superfield}
  \Phi^\text{(SDBI)}(\eta) \,=\, \gamma^+ + \eta^A \psi_A
  + {1\over 2!} \eta^{A}\eta^{B}S_{AB}
  + {1\over 3!} \eta^{A}\eta^{B}\eta^{C} \epsilon_{ABCD} \bar\psi^{D}
  + \eta^{1}\eta^{2}\eta^{3}\eta^{4}\, \gamma^-,
\end{align}
where the supermultiplet contains photons, photinos and scalars. It is well known that for photon scatterings in Born-Infeld theory, only helicity-conserved amplitudes with even multiplicity are non-vanishing. By supersymmetry this generalizes to the superamplitude, thus we will only have the middle sector $k=n/2$ for even $n$. We omit the subscript $k=n/2$ of the measure, and write
\begin{align}\label{DBImeasure}
  d\mu^{(\cal N)}_{n} := d\mu^{(\cal N)}_{n,\frac n 2}
  = {d^{2n}\sigma \over \operatorname{vol} {\rm GL}(2,{\mathbb C})}
  \sum_{I=1}^{n/2} \delta^{2|\cal N} \left( \{\tilde\lambda_I|\eta_I\}
  - \sum_{i=n/2+1}^n {\{\tilde\lambda_i|\eta_i\}\over (I\,i)} \right)
  \sum_{i=n/2 +1}^{n} \delta^{2} \left( \lambda_i  -  \sum_{I=1}^{n/2} {\lambda_I \over (i\,I)} \right).
\end{align}
Recall that it is permutation invariant and identical to that with the polynomial form equations.

It turns out that we only need one more ingredient for writing down the formulas for amplitudes in all the three theories. We define an $n\times n$ antisymmetric matrix $A_n$ with entries $A_{ab} = {s_{ab} \over (a\,b)}$ for $a \ne b$ and $A_{aa} = 0$. It has two null vectors and we define the reduced pfaffian and determinant as
\begin{align}\label{Det-A}
  {\rm Pf}' A_n \,:=\, {(-)^{a{+}b} \over (a\,b)}\, {\rm Pf}\, |A|^{a,b}_{a,b}\,,
  \quad {\det}'A_n \,=\, \big({\rm Pf}' A_n\big)^2\,.
\end{align}
One can show that the rank of the matrix $A_n$ is less than $n{-}2$ when we plug in the solutions of 4d scattering equations in any sector except the middle sector $k = n/2$~\cite{Cachazo-He-Yuan-1412.3479}.
Thus ${\det}'A_n$ is only non-vanishing for the sector $k=n/2$, which already suggests strongly that it should appear in the formula for SDBI.  The formula for the complete tree-level S-matrix in ${\cal N}\!=\!4$ super-DBI reads:
\begin{empheq}[box=\fbox]{align}
  \label{SDBI}
  {\cal M}^\text{SDBI}_{n}
  \,=\, \int d\mu^{(4)}_{n}\, {\det}'A_{n}.
\end{empheq}

As shown in~\cite{Cachazo-He-Yuan-1412.3479}, we have double-copy relations for special Galileon theory and super-DBI:
\begin{align}\label{KLT}
  {\rm sGal} \,=\, {\rm NLSM} \otimes {\rm NLSM},\quad
  {\rm BI} \,=\, {\rm YM} \otimes {\rm NLSM}, \quad
  {\rm SDBI} \,=\, {\rm SYM} \otimes {\rm NLSM},
\end{align}
where the last relation follow from the second one by supersymmetry. From these relations, it has become clear that the formula for NLSM and sGal must take the form
\begin{empheq}[box=\fbox]{align}
  \label{NLSM1}
  &{\cal M}^\text{NLSM}_{n}(1,2,\ldots, n)
  \,=\, \int d\mu^{(0)}_{n} {1\over (12)(23) \cdots (n1)}
  {{\det}' A_n \over {H}_{n}},\\
  &\label{sGal1}
  {\cal M}^\text{sGal}_{n}
  \,=\, \int d\mu^{(0)}_{n}\, {\big( {\det}' A_n \big)^2 \over {H}_{n}},
\end{empheq}
where we have defined ${H}_n := {\det}'{\mathbb H}_{n/2}\, {\det}'\overline{\mathbb H}_{n/2}$.  Unlike the bi-adjoint $\phi^3$ theory, these scalar amplitudes are only non-vanishing for the $k=n/2$ sector of the solutions to 4d scattering equations. This can be explained from the appearance of ${\det}'A_n$, as already noticed in~\cite{Cachazo-He-Yuan-1412.3479}. The double-copy relations \eqref{KLT} also specify to the middle sector in 4d, where only the term $k=n/2$ in \eqref{phi3} is needed~\cite{Cachazo-Zhang-1601.06305}.

There is a further relation which makes these formulas much simpler than (the middle-sector) $\phi^3$ amplitudes. As we checked up to ten points, $H_n$ and ${\rm Pf}' A_n$ are actually proportional to each other:
\begin{align}
{\rm Pf}' A_n \,=\, \det J_{n \over 2}\,H_n\,,
\end{align}
where the proportionality factor is $\det J_{n \over 2}$ with entries of the matrix of the form $(I\,i)^{-1}$ for rows labelled by $I=1,\ldots, {n/2}$ and columns by $i={n/2}+1,\ldots, n$. It is straightforward to find
\begin{align}\label{}
  \det J_{n\over 2} \,=\,
  {\prod_{J<K}(J\,K)\, \prod_{j<k}(j\,k) \over \prod_{I,i} (I\,i)}.
\end{align}
We will not prove this very interesting identity in the paper, but just to say that it simplifies \eqref{NLSM1} and \eqref{sGal1} further:
\begin{empheq}[box=\fbox]{align}
  \label{NLSM}
  &{\cal M}^\text{NLSM}_{n}(1,2,\ldots, n)
  \,=\,\int d\mu^{(0)}_{n}
  {1\over (12)(23) \cdots (n1)}\,\det J_{n\over 2}\,{\rm Pf}' A_n ,\\
  &\label{sGal}
  {\cal M}^\text{sGal}_{n}
  \,=\, \int d\mu^{(0)}_{n}\, \det J_{n\over 2}\, \big( {\rm Pf}' A_n \big)^3.
\end{empheq}
In this form, it becomes very clear that, unlike the $\phi^3$ case, there is no spurious pole for amplitudes in NLSM or sGal, and their formulas take a much simpler form. For NLSM, DBI and sGal, their formulas contain $({\rm Pf}' A_n)^t$ for $t=1,2,3$ respectively.


We have very strong evidence for the new 4d formulas, \eqref{SDBI}, \eqref{NLSM}, \eqref{sGal}, by comparing with their general-dimension CHY formulas, or by studying their factorization properties directly. More explicitly, we have computed numerically up to six points and verify that they give correct amplitudes. For example, by directly evaluating \eqref{SDBI} for $n=4$ we find
\begin{align}\label{}
  {\cal M}^\text{${\cal N} \!=\! 4$ SDBI}_{4}
  \,=\, \delta^{4}(P)\, \delta^{0|8}(Q)\,
  \frac{[3\,4]^2}{\braket{1\,2}^2}\,.
\end{align}
Similarly we have checked six-scalar amplitudes in all three theories, and in ${\cal N}\!=\!4$ super-DBI six-photon amplitudes \cite{Boels-Larsen-Obers-Vonk-0808.2598}, two-fermions-four-photon and two-scalar-four-photon amplitudes \cite{Chen-Huang-Wen-1505.07093}, as well as six-fermion amplitudes\footnote{The six-fermion amplitude in Volkov-Akulov theory was obtained in \cite{Luo:2015tat}, and very recently reproduced in \cite{Cachazo:2016njl} using a formula similar to our~\eqref{SDBI}.}.

\section{Double soft theorems}\label{sec3}

In this section, as both consistency checks and more importantly applications of the new 4d formulas proposed in the previous section, we derive the double soft theorems in ${\cal N} \!=\! 4$ super-DBI, NLSM, sGal. We also discuss some double limits in ${\cal N}\!=\!4$ SYM and ${\cal N}\!=\!8$ SUGRA~\cite{Geyer-Lipstein-Mason-1404.6219}.

As shown in \cite{Cachazo-He-Yuan-1503.04816}, in the simultaneous double soft limit, there are two types of solutions to the scattering equations\,--\,those non-degenerate ones, i.e.~all $\sigma$'s are distinct from each other, and a {\it unique degenerate} solution with the two $\sigma$'s of the soft legs coincide. We find exactly the same conclusion for the solutions of 4d scattering equations \eqref{4dr}.

The key observation~\cite{Cachazo-He-Yuan-1503.04816} is that, when the contribution of the degenerate solution dominates over that of  non-degenerate ones in the double soft limit, one can derive double soft theorems by evaluating the formula for the degenerate solution only.  Here we will see that it is indeed the case for all super-amplitudes in ${\cal N}\!=\!4$ super-DBI involving the emission of a pair of soft photons, fermions or scalars.

\subsection{Double soft theorems in ${\cal N} \!=\! 4$ super-DBI}

Let us start with an $(n+2)$-point amplitude with even $n$ in ${\cal N} \!=\! 4$ super-DBI theory,
\begin{align}\label{}
  {\cal M}_{n+2} \,=\, \int d\mu^{(4)}_{n+2}\, {\det}' A_{n+2},
\end{align}
and here we write the measure $d\mu_{n+2}^{(4)}$ as,
\begin{align}\label{}
  &{d^{2(n+2)}\sigma \over \operatorname{vol} {\rm GL}(2,{\mathbb C})}
  \prod_{I=1}^{n/2} \delta^{2|4}\!
  \left(\{\tilde\lambda_I|\eta_I\} - \sum_{i=n/2+1}^{n}{\{\tilde\lambda_i|\eta_i\} \over (I\,i)}
  - {\{\tilde\lambda_p|\eta_p\} \over (I\,p)}\right)
  \delta^{2|4}\!\left(\{\tilde\lambda_{q}|\eta_q\}
  - \sum_{i=n/2+1}^{n}{\{\tilde\lambda_i|\eta_i\} \over (q\,i)}
  - {\{\tilde\lambda_p|\eta_p\} \over (q\,p)}\right)
  \nonumber\\
  &\times
  \prod_{i=n/2+1}^n \delta^{2}
  \left(\lambda_i - \sum_{I=1}^{n/2} {\lambda_I \over (i\,I)}
  - {\lambda_{q} \over (i\,q)} \right)
   \delta^{2}
  \left(\lambda_{p} - \sum_{I=1}^{n/2} {\lambda_I \over (p\,I)}
  - {\lambda_{q} \over (p\,q)}\right),
  \label{N=4-measure-n+2}
\end{align}
where $I=n+2, 1, \ldots, n/2$ and $i=n/2+1, \ldots, n, n+1$. For the sake of brevity, here and in the rest of this paper we denote the indices $n+1$ and $n+2$ as $p$ and $q$ respectively.

To be concrete, we perform anti-holomorphic and holomorphic soft limits for the external legs $p$ and $q$ respectively, and introduce a small real parameter $\epsilon$ to control this simultaneous double soft limit:
\begin{align}\label{soft-limit}
  \tilde\lambda_p \,\rightarrow\, \epsilon\tilde\lambda_p, \quad
  \lambda_q \,\rightarrow\, \epsilon\lambda_q,
\end{align}
while $\lambda_p, \eta_p$ and $\tilde\lambda_q, \eta_{q}$ stay finite \cite{Arkani-Hamed-Cachazo-Kaplan-0808.1446}. In this limit, we have $(a\,b)\sim {\cal O}(1)$ for non-degenerate solutions, while for the degenerate solution,  $(p\,q)\sim {\cal O}(\epsilon)$.

Now we can study the scaling behavior of the formula in $\epsilon$ for both degenerate and non-degenerate solutions. In the double soft limit \eqref{soft-limit}, the bosonic part of measure \eqref{N=4-measure-n+2} behaves as $d\mu_{n+2}^{(0)} \sim {\cal O}(\epsilon)$ for the degenerate solution and $d\mu_{n+2}^{(0)} \sim {\cal O}(\epsilon^0)$ for non-degenerate solutions while ${\det}' A_{n+2}\sim{\cal O}(\epsilon^4)$  and
${\det}' A_{n+2}\sim{\cal O}(\epsilon^2)$ for the degenerate solution and non-degenerate ones respectively.

We also need to consider the scaling behavior from fermionic delta functions in the measure \eqref{N=4-measure-n+2}, which strongly depends on the SU(4) flavors of the soft particles. Let first recall the on-shell superfield \eqref{N=4-on-shell-superfield} and the following fermionic $\delta$-function in the measure \eqref{N=4-measure-n+2}
\begin{align}\label{eta-delta-fun-part-n+2}
  \prod_{I=1}^{n/2} \delta^{0|4}
  \left(\eta_I - \sum_{i=n/2+1}^{n}{\eta_i \over (I\,i)}
  - {\eta_p \over (I\,p)}\right) \times  \delta^{0|4}\left(\eta_{q} - \sum_{i=n/2+1}^{n}{\eta_i\over (q\,i)}
  - {\eta_p\over (q\,p)}\right).
\end{align}
While it is obvious that for any pair of soft particles, it is ${\cal O}(1)$ for non-degenerate solutions in the limit \eqref{soft-limit}, the case for the degenerate solution is more subtle. One needs to distinguish between two cases:~(i) when the two soft particles form a SU(4) flavor-singlet, i.e.~$(\gamma^+,\, \gamma^-)$ photon pair, $(\psi_A,\, \bar\psi^A)$ fermion pair, or $(S_{AB},\, S^{AB})$ scalar pair, and (ii) when they do not form a singlet, e.g.~$(\psi_A, \bar\psi^B)$ or $(S_{AD}, \, S^{BD})$.

For the first case, the leading-order contribution comes from picking out all $\eta_p, \eta_q$ from the last fermonic delta function of \eqref{eta-delta-fun-part-n+2}, and the remainder becomes exactly fermionic delta functions for $n$-point formula. The last fermonic delta function evaluates to $1/(p\,q)^{2-2s}$  which behaves as $\sim {\cal O}(\epsilon^{2s-2})$, where ``$s$''  denotes the spin of the soft pair. For the second case, we also have one $\eta_p$ from other fermonic delta functions, and the factor becomes $1/(p\,q)^{1-2s}$. When combining with the bosonic measure and integrand, for both cases the contribution from degenerate solution always dominates.

The second case is sub-leading compared to the first case, so we refer to the latter as the ``leading-order" double-soft theorems and the former as the ``sub-leading" ones. We first discuss the leading-order case, and postpone the very interesting discussion of the subleading case to the end of this subsection.

It is convenient to introduce the change of variable for the degenerate solution~\cite{Cachazo-He-Yuan-1503.04816}
\begin{align}\label{variable-transf-xi-rho}
  \sigma_p \,=\, \rho - \epsilon\,{\xi\over 2}, \quad
  \sigma_q \,=\, \rho + \epsilon\,{\xi\over 2}
\end{align}
with $\sigma_{q p}=\epsilon\,\xi\sim {\cal O}(\epsilon)$, and we have $d\sigma_p d\sigma_q  \,=\, \epsilon\, d\rho\, d\xi$. In these variables, the integrand, $\det{}' A$, becomes
\begin{align}
  \label{det-A-soft-exapnsion}
  {\det}' A_{n+2} \,=\, \epsilon^2\, {s_{pq}^2 t_p^2t_q^2 \over \xi^2} {\det}' A_{n} + {\cal O}(\epsilon^4),
\end{align}
and we can write the complete measure involving a pair of soft particles of spin $s$ in a unified form:
{\small \begin{align}
  \epsilon\,\left(-{\epsilon \xi \over t_pt_q}\right)^{-2(1-s)}\,
  {dt_p dt_q \over t_p^3t_q^3}\,
  d\rho\,d \xi \,
  &\delta^{2}\!\left(\tilde\lambda_{q}
  - \sum_{i=n/2+1}^{n}{\tilde\lambda_i\over (q\,i)}
  - {t_pt_q\tilde\lambda_p\over \xi}\right)~\delta^{2}\!\left(\lambda_{p} - \sum_{I=1}^{n/2} {\lambda_I \over (p\,I)}
  + {t_pt_q\lambda_{q} \over \xi}\right)~d\mu^{(4)}_n  + {\cal O}(\epsilon^{2s}).\nonumber
\end{align}}

Our task is to perform the integral over~$t_p, t_q, \xi$ and $\rho$~by using the four additional delta functions above. For this purpose it is convenient to rewrite these delta functions as
\begin{align}\label{delta-fun-sim}
\begin{aligned}
  \delta^{2}\! \left(\tilde\lambda_{q} - \!\!\sum_{i=n/2+1}^{n}{\tilde\lambda_i\over (q\,i)}
  - {t_pt_q\tilde\lambda_p \over \xi} \right)
  \,&=\,
  {1\over t_pt_q [pq]}
  \delta\! \left( 1 - \!\!\sum_{i=n/2+1}^{n} {[p\,i] \over [pq]} {t_qt_i\over\sigma_{qi} } \right)
  \delta\! \left( \sum_{i=n/2+1}^{n} {[qi] \over [qp]} {t_i\over t_p\sigma_{qi}}
  + {1 \over \xi} \right),
  \\
  \delta^{2}\! \left(\lambda_{p} - \sum_{I=1}^{n/2} {\lambda_I \over (p\,I)}
  + {t_pt_q\lambda_q \over \xi}\right)
  \,&=\,
  {-1\over t_pt_q \braket{pq}}
  \delta\!
  \left(1 - \sum_{I=1}^{n/2} {\braket{qI} \over \braket{qp}} {t_pt_I\over \sigma_{pI} } \right)
  \delta\! \left( \sum_{I=1}^{n/2} {\braket{pI} \over \braket{pq}} {t_I\over t_q\sigma_{pI} }
  - {1 \over \xi} \right).
\end{aligned}
\end{align}
It is clear now that from the RHS of \eqref{delta-fun-sim}, we can use the two delta functions without $\xi$ to fix $t_p, t_q$:
\begin{align}\label{t_p-t_q}
  t_p^{-1} \,=\, \sum_{I=1}^{n/2} {\braket{qI} \over \braket{qp}} {t_I\over \sigma_{pI}},
  \qquad
  t_q^{-1} \,&=\, \sum_{i=n/2+1}^{n} {[pi] \over [pq]} {t_i\over \sigma_{qi}}.
\end{align}
After integrating out $t_p, t_q$, the formula in the double soft limit \eqref{soft-limit} becomes
\begin{align}
  {\cal M}^{(s)}_{n+2}
  \,&=\, (-1)^{1-2s}\epsilon^{1+2s} \int d\mu_n^{(4)} {\det}' A_n
  \int d\rho d\xi\,
  {s_{pq} \over (t_p t_q)^{2s}\, \xi^{4-2s}}
  \delta(f_1)\delta(f_2)
  + {\cal O}(\epsilon^{2+2s}),
  \label{n+2-pt-amp-soft-limit-X-1}
\end{align}
where we used the superscript $(s)$ for the spin of the soft pair. Here we also denote
\begin{align}
  f_1 \,=\, \sum_{i=n/2+1}^{n} {[qi] \over [qp]} {t_i\over t_p} {1\over \sigma_{qi}}
  + {1\over \xi}
  \,&=\, -{1\over s_{pq}} \sum_{i=n/2+1}^{n}\sum_{I=1}^{n/2}
  {[i|q\!\ket{I} t_It_i\over \sigma_{pI}\sigma_{qi}}
  + {1\over \xi},
  \\
  f_2 \,=\, \sum_{I=1}^{n/2} {\braket{pI} \over \braket{pq}}
  {t_I\over t_q} {1\over\sigma_{pI}} - {1\over \xi}
  \,&=\, -{1\over s_{pq}} \sum_{i=n/2+1}^{n}\sum_{I=1}^{n/2}
  {[i|p\!\ket{I} t_It_i\over \sigma_{pI}\sigma_{qi}}
  -  {1\over \xi}\,,
\end{align}
and in the second equality we have plugged in the solution for $t_p, t_q$.

Now the problem of integrating over $\rho$ and $\xi$ resembles that in deriving double soft theorems in arbitrary dimensions in \cite{Cachazo-He-Yuan-1503.04816}, and we recall the transformation of the delta functions,
\begin{align}\label{delta-fun-fp-fm}
   \delta(f_1) \delta(f_2) \,=\, -2\delta(f_1+f_2) \delta(f_1-f_2)\,.
\end{align}
The key point here is to note that $f_1\pm f_2$ can be simplified to particularly nice form as a sum over $\{1,\ldots,n\}$\,!
Let us make a partial fraction decomposition for $1/\sigma_{pI}\sigma_{qi}$, then $f_2+f_2$ can be written as
\begin{align}
  f_1+f_2 \,&=\,- {1 \over s_{pq}}
  \sum_{i=n/2+1}^n\sum_{I=1}^{n/2}
  \left( {1 \over \rho-\sigma_i} - {1 \over \rho-\sigma_I} \right)
  {[i| (p+q) \ket{I} \over (i\,I) } \nonumber\\
  \,&=\,- {1 \over s_{pq}}\left\{
  \sum_{i=n/2+1}^n {1 \over \rho-\sigma_i}
  \sum_{I=1}^{n/2} {[i| (p+q) \ket{I} \over (i\,I)}  +
  \sum_{I=1}^{n/2} {1 \over \rho-\sigma_I}
  \sum_{i=n/2+1}^n {[i| (p+q) \ket{I} \over (I\,i) }
  \right\}\,.
\end{align}
By 4d scattering equations \eqref{4dr}, the two inner sums simply give  $[ i | p+q | i \rangle$ and $[I | p+q | I \rangle$, and
\begin{align}
  f_1+f_2
  \,=\, {1 \over s_{pq}}
  \sum_{a=1}^n {2 k_a\cdot (p+q) \over \rho-\sigma_a}.
  \label{f_1+f_2-final}
\end{align}
The same technique works for $f_1-f_2$, and one obtains immediately the solution for $\xi$ from $f_1-f_2=0$ as follows
\begin{align}\label{}
  \xi^{-1}
  \,&=\, {1 \over 2s_{pq}}
  \sum_{i=n/2+1}^n\sum_{I=1}^{n/2}
  \left( {1 \over \rho-\sigma_i} - {1 \over \rho-\sigma_I} \right)
  {[i| (p-q)\ket{I} \over (I\,i)}
  \,=\, {1 \over s_{pq}}
  \sum_{a=1}^n {k_a\cdot (p-q) \over \rho-\sigma_a}.
  \label{f_1-f_2-xi-final}
\end{align}
By the way, from eq.~\eqref{t_p-t_q} and eq.~\eqref{variable-transf-xi-rho} one can get a similar result for $t_p t_q$:
\begin{align}
  t_p^{-1}t_q^{-1} \,&=\, {1\over s_{pq}}
  \sum_{i=n/2+1}^n \sum_{I=1}^{n/2}
  \left( {1 \over \rho-\sigma_i} - {1 \over \rho-\sigma_I} \right)
  {[pi]\braket{Iq} \over (i\,I)}
  \,=\, {1\over s_{pq}}
  \sum_{a=1}^n {[pa]\braket{aq} \over \rho-\sigma_a}.
  \label{t_p-t_q-final}
\end{align}

Now we can package everything together. First we localize the $\xi$-integral by $\delta(f_1 - f_2)$, and regard the $\rho$-integral
as a contour integral with contour ${\cal C}$ encircling the zeroes of $f_1+f_2=0$,
\begin{align}
  {\cal M}^{(s)}_{n+2} \,&=\, (-1)^{1-2s} \epsilon^{1+2s} \int d\mu_n^{(4)} {\det}' A_n
  \oint_{\cal C} {d\rho \over 2\pi i}
  {s_{pq}\, (t_p t_q)^{-2s}\, \xi^{-2(1-s)} \over f_1+f_2}
  + {\cal O}(\epsilon^{2+2s})\,.
  \label{n+2-pt-amp-soft-limit-X-2}
\end{align}
Plugging eqs.~\eqref{f_1+f_2-final},~\eqref{f_1-f_2-xi-final},~\eqref{t_p-t_q-final}~into eq.~\eqref{n+2-pt-amp-soft-limit-X-2} immediately gives
\begin{align}
  {\cal M}^{(s)}_{n+2} \,&=\, (-\epsilon)^{1+2s}
  \int d\mu_n^{(4)} {\det}' A_n
  \oint_{\cal C} {d\rho \over 2\pi i}
  {\left(\sum_{a=1}^n {[p|a\ket{q} \over \rho-\sigma_a}\right)^{2s}
  \left(\sum_{b=1}^n {k_b\cdot (p-q) \over \rho-\sigma_b}\right)^{2(1-s)}
  \over
  \sum_{c=1}^n {k_c \cdot (p+q) \over \rho-\sigma_c}
  }
  + {\cal O}(\epsilon^{2+2s}).
  \nonumber
\end{align}
This integral do not receive the contribution from a simple pole at $\rho=\infty$ due to momentum conservation in the numerator.
Thus we only need to consider simple poles at $\rho=\sigma_a$ with $a=1,2,\ldots, n$ and obtain by the residue theorem
\begin{empheq}[box=\fbox]{align}
  {\cal M}^{(s)}_{n+2} \,&=\, \epsilon^{1+2s}
  \sum_{a=1}^n {
  \big( k_a \cdot (q-p) \big)^{2-2s} [p|a\!\ket{q}^{2s} \over 2k_a \cdot (p+q)}
  {\cal M}^{(s)}_n
  + {\cal O}(\epsilon^{2+2s}).
  \label{Double-soft-super-DBI-singlet}
\end{empheq}
It is highly non-trivial that the combinations appeared, $f_1+ f_2$, $f_1-f_2$ and $t_p t_q$, all become a sum over $a=1,\ldots,n$, which is what we need to derive the nice soft theorems~\eqref{Double-soft-super-DBI-singlet}. The key for this to happen is the use of scattering equations \eqref{4dr}. Note that these theorems now directly hold for superamplitudes in four dimensions, i.e.~hard particles can be any particles in supermultiplet \eqref{N=4-on-shell-superfield}.

The double soft photon limit ($s=1$) and double soft scalar limit ($s=0$) in the DBI theory are obtained using CHY representations in \cite{Cachazo-He-Yuan-1503.04816}, while the double fermion limit for $s={1\over 2}$ without flavors is conjectured by studying six-fermion amplitudes in Volkov-Akulov theory \cite{Chen-Huang-Wen-1412.1809}. Here we have shown that these seemingly different double soft theorems can be unified for superamplitudes in ${\cal N}\!=\!4$ SDBI, and this unified form \eqref{Double-soft-super-DBI-singlet} certainly deserves further study.

\subsubsection{Sub-leading theorems in ${\cal N}=4$ super-DBI}

Now we turn to the case that the two soft particles are not in a flavor singlet of SU(4), and for simplicity we consider $(\psi_A, \bar\psi^B)$ fermion-pair, and $(S_{AD}, S^{BD})$ scalar-pair.

For convenience, let us first rewrite the fermionic $\delta$-function \eqref{eta-delta-fun-part-n+2} here
\begin{align}
  \prod_{I=1}^{n/2} \delta^{0|4}
  \left(\eta_I - \sum_{i=n/2+1}^{n}{\eta_i \over (I\,i)}
  - {\eta_p \over (I\,p)}\right) \times  \delta^{0|4}\left(\eta_{q} - \sum_{i=n/2+1}^{n}{\eta_i\over (q\,i)}
  - {\eta_p\over (q\,p)}\right),
  \nonumber
\end{align}
and take a closer look. Unlike the single-flavor case, here we pick $\eta_p^A$ from one of those $\delta$-functions with $\eta_I$, and the remaining three $\eta$'s, $(\eta^3_q)_B$ for $s={1\over 2}$ or $\eta_p^D (\eta^2_q)_{B D}$ for $s=0$, from the last $\delta$-function. The operation of extracting $\eta_p^A$ from those $\delta$-functions amounts to taking derivative $\partial/\partial\eta_I$ with a factor $1/(I\,p)$ and a sum over $I$. Furthermore, an additional $\eta$ from the last $\delta$-function must come from the sum $\sum_i \eta_i/(q\,i)$.
To be more precise, by projecting upon the relevant terms in the $\eta_p$ and $\eta_q$, one finds the fermionic part of the measure contributing to the leading soft limits,
\begin{align}
  &- \left(-{\epsilon\xi \over t_pt_q}\right)^{2s-1} \sum_{i=n/2+1}^n \sum_{I=1}^{n/2} {1\over (q\,i)(I\,p)}\eta_i^B {\partial \over \partial\eta_I^A}
  \,\delta^{(2n)}\!\big({\cal F}_n\big)
  + {\cal O}(\epsilon^{2s})
  \nonumber\\
  \,&=\, -(-\epsilon)^{2s-1}
  {(t_pt_q)^{2(1-s)} \over \xi^{1-2s}}
  \left\{
  \sum_{I=1}^{n/2} {1\over\rho-\sigma_I}
  \sum_{i=n/2+1}^n {\eta_i^B {\partial_{\eta_I^A}} \over (i\,I)}
  +
  \sum_{i=n/2+1}^n {1\over\rho-\sigma_i}
  \sum_{I=1}^{n/2} {\eta_i^B {\partial_{\eta_I^A}} \over (I\,i)}
  \right\}
  \delta^{(2n)}\!\big({\cal F}_n\big) + {\cal O}(\epsilon^{2s})
  \nonumber\\
  \,&=\, -(-\epsilon)^{2s-1}\,
  {(t_pt_q)^{2(1-s)} \over \xi^{1-2s}}
  \sum_{a=1}^n {(R_a)^B_{~A}\over \rho-\sigma_a}
  \,\delta^{(2n)}\!\big({\cal F}_n\big)
  + {\cal O}(\epsilon^{2s}),
  \label{fermionic-part-soft-limit}
\end{align}
where we denote the product of fermionic $\delta$-functions, and the SU(4) generator on the leg $a$ as:
\begin{align}\label{}
  \delta^{(2n)}\big({\cal F}_n\big)
  \,\equiv\, \prod_{I=1}^{n/2} \delta^{0|4}
  \left(\eta_I - \sum_{i=n/2+1}^{n}{\eta_i \over (I\,i)} \right),\qquad (R_a)_{~A}^B \,\equiv\, \eta_a^B {\partial\over\partial\eta_a^A}\,.
\end{align}
In eq.~\eqref{fermionic-part-soft-limit}, we have used the same trick as the bosonic case, and the fermionic scattering equations:
\begin{align}
  \sum_{i=n/2+1}^n {1 \over (I\,i)} \eta_i^A \,=\, \eta_I^A,
  \qquad
  \sum_{I=1}^{n/2} {1 \over (i\,I)} {\partial\over \partial\eta_I^A}
  \,=\, {\partial\over \partial\eta_i^A},
\end{align}
where the second set of equations follow from those equations for $\tilde\eta$'s (written in the conjugate superspace) by the replacement $\tilde\eta \to \partial/\partial\eta$.

Note $\text{eq.~\eqref{fermionic-part-soft-limit}}\sim{\cal O}(\epsilon^{2s-1})$ as we claimed, which means that the double-soft behavior is sub-leading for non-singlet soft pair, compared to the singlet pair. However, recall behavior of $d\mu^{(0)}$ and $\det{}' A$, it is still the case that the degenerate solution is dominant at this order, see also table \ref{Soft-Limits-Amp-table} at the end of the section. By eq.~\eqref{fermionic-part-soft-limit} and repeating the exact same derivation gives for ${\cal M}^{(s)}_{n+2}$
\begin{align}
  -(-\epsilon)^{2+2s} s_{pq}  \int d\mu^{(0)}_n {\det}' A_n
  \oint_{\cal C} {d\rho \over 2\pi i}
  {\left(\sum\limits_{a=1}^n {[p|a\ket{q} \over \rho-\sigma_a}\right)^{2s}
  \left(\sum\limits_{b=1}^n {k_b\cdot (p-q) \over \rho-\sigma_b}\right)^{1-2s}
  \sum\limits_{c=1}^n {(R_c)_{~A}^B \over \rho-\sigma_c}
  \delta^{(2n)}\!\big({\cal F}_n\big)
  \over
  \sum\limits_{d=1}^n {k_d \cdot (p+q) \over \rho-\sigma_d}
  }
  + {\cal O}(\epsilon^{3+2s}).
  \nonumber
\end{align}
Similarly, performing the $\rho$-integral by encountering simple poles at $\rho=\sigma_a$ yields
\begin{empheq}[box=\fbox]{align}
  {\cal M}^{(s)}_{n+2}
  \,&=\,
  -\epsilon^{2+2s} s_{pq}
  \sum_{a=1}^n
  { \left(k_a\cdot (q-p)\right)^{1-2s} [p|a\!\ket{q}^{2s} \over 2k_a\cdot (p+q) }
  \eta_a^B {\partial\over\partial\eta_a^A}
  {\cal M}^{(s)}_n
  + {\cal O}(\epsilon^{3+2s})
\end{empheq}
for two soft fermions $(\psi_A, \bar\psi^B)$ emission ($s={1\over 2}$) and two soft scalars $(S_{AD}, S^{BD})$ emission ($s=0$) respectively. The result bears striking similarity with the double soft scalar theorem in ${\cal N}\!=\!8$ SUGRA \cite{Arkani-Hamed-Cachazo-Kaplan-0808.1446} (see \cite{Volovich-Wen-Zlotnikov-1504.05559,Klose-McLoughlin-Nandan-Plefka-Travaglini-1504.05558, Vecchia-Marotta-Mojaza-1507.00938} for recent works on double soft behavior in~${\cal N}\!=\!4$ SYM).
In that case, the theorem directly probes the coset structure ($E_{7(7)}/{\rm SU}(8)$) of the vacua, and we hope that our results here, which has similar structure, can be useful for studying the coset structure of ${\cal N}\!=\!4$ SDBI theory.

\subsection{More double-soft theorems}
Having established all double-soft theorems in super-DBI, we now briefly discuss double soft theorems for NLSM, sGal, as well as those in ${\cal N}\!=\!4$ SYM and ${\cal N}\!=\!8$ SUGRA.
For color-ordered amplitudes in SYM and NLSM, we will focus on the case that the soft particles are adjacent.

All we need are the behavior of the Parke-Taylor factor and that for ${\det}'\mathbb{H}\, {\det}'\overline{\mathbb H}$, in the double soft limit.
 For non-degenerate solutions, the Parke-Taylor factor has leading order behavior of ${\cal O}(1)$, while for the degenerate solution, it is straightforward to get
\begin{align}\label{}
  {1\over (12) \cdots (np)(pq)(q1)}
  \,=\, {1\over (12) \cdots (n1)}
  {t_p^2 t_q^2 \over \epsilon\,\xi}
  \left( {1\over\rho-\sigma_n} - {1\over\rho-\sigma_1} \right)
  + {\cal O}(\epsilon^0),
\end{align}
in the double soft limit \eqref{soft-limit}.
Similarly, in the double limit, ${\det}'\mathbb{H}\, {\det}'\overline{\mathbb H} \sim {\cal O}(\epsilon^2)$ for non-degenerate solutions, while for degenerate solution, we have
\begin{align}\label{}
  {\det}' \mathbb{H}_{k+1} {\det}' \overline{\mathbb H}_{n-k+1}
  \,&=\, \epsilon^2\, \left(-\sum_{I=1}^k \mathbb{H}_{qI}\right){\det}'\mathbb{H}_{k}
  \left(-\sum_{i=k+1}^n \mathbb{\mathbb H}_{pi}\right){\det}'\overline{\mathbb H}_{n-k}
  + {\cal O}(\epsilon^3)
  \\\nonumber
  \,&=\, - \epsilon^2\, t_pt_q\sum_{a=1}^n {[p|a\ket{q} \over \rho-\sigma_a}
  {\det}'\mathbb{H}_{k}\, {\det}'\overline{\mathbb H}_{n-k}
  + {\cal O}(\epsilon^3)
  \\\nonumber
  \,&=\, -\epsilon^2\, s_{pq}\,
  {\det}'\mathbb{H}_{k}\, {\det}'\overline{\mathbb H}_{n-k}
  + {\cal O}(\epsilon^3),
\end{align}
where the same trick as the case for $f_1 \pm f_2$ and $t_pt_q$ is nicely used again.
Of course, it also holds for $k=n/2$, namely $H_{n+2} = -\epsilon^2 s_{pq} H_n + {\cal O}(\epsilon^3)$ in the same limit.

We summarize the soft scaling behavior in $\epsilon$ for all the (bosonic) building blocks in table \ref{Soft-Limits-Blocks-table}.
\begin{table}[!h]
\centering
\begin{tabular}{ccc}
\hline\hline
Building Block & ~~${\cal O}(\text{d})$~~ & ~~${\cal O}(\text{nd})$~~ \\ [0.1ex]
\hline
$d\mu^{(0)}$   &  1  & 0 \\
${\det}'A$     & $2$ & 4 \\
Parke-Taylor factor   & -1  & 0 \\
${\det}'{\mathbb H}\,{\det}'\overline{\mathbb H}$  & 2  & 2
\\ [0.1ex]
\hline\hline
\end{tabular}
\caption{Leading scaling behavior in soft parameter $\epsilon$ of the building blocks in the limit \eqref{soft-limit}. Here ``d'' and ``nd'' stand for the degenerate and non-degenerate solutions respectively.}
\label{Soft-Limits-Blocks-table}
\end{table}

For U$(N)$ NLSM and the special Galileon theory, let us recall the formula for their amplitudes:
\begin{align}\label{}
  {\cal M}^\text{NLSM}_{n+2}
  \,=\, \int d\mu^{(0)}_{n+2}\,
  {1\over (12) \cdots (np)(pq)(q1)}
  {{\det}' A_{n+2} \over H_{n+2}},\qquad
  {\cal M}^\text{sGal}_{n+2}
  \,=\, \int d\mu^{(0)}_{n+2}\,
  {\big({\det}' A_{n+2}\big)^2 \over H_{n+2}}.
\end{align}
By power counting of the soft parameter $\epsilon$ for building blocks, again we find the soft scalar limits at leading order only receive the contribution from the degenerate solution. The same derivation as for SDBI gives the leading double soft scalar theorems:
\begin{align}\label{}
  {\cal M}_{n+2}(1,\ldots,n,p,q) \,=\, \epsilon^m {\cal S}\, {\cal M}_n(1,\ldots,n)
  + {\cal O}(\epsilon^{m+1}),
\end{align}
where $m=0$ for NLSM and $m=3$ for sGal, and soft factors are given respectively by
\begin{align}\label{}
  {\cal S}_{\rm NLSM} \,=\,
  {k_n \cdot (p-q) \over 2k_n \cdot (p+q)} + {k_1 \cdot (q-p) \over 2k_1\cdot(q+p)},\qquad  {\cal S}_{\rm sGal} \,=\,
  s_{pq} \sum_{a=1}^n {\big( k_a \cdot (p-q) \big)^2 \over 2k_a\cdot(p+q)}\,,
\end{align}
which coincide with the leading-order results of \cite{Cachazo-He-Yuan-1503.04816}. Note that single and double scalar emissions in NLSM were also investigated in \cite{Kampf-Novotny-Trnka-1304.3048, Du-Luo-1505.04411, Low-1512.01232}.

Finally we make a classification of double soft theorems for ${\cal N}\!=\!4$ SYM and ${\cal N}\!=\!8$ SUGRA. Unlike the case for the other three theories, the degenerate solution does not always dominate for leading double soft limit in ${\cal N}\!=\!4$ SYM and ${\cal N}\!=\!8$ SUGRA, as listed in table \ref{Soft-Limits-Amp-table}. For SYM, the degenerate solution still dominates for the following three cases, giving double-soft theorems:
\begin{align}
  {\cal M}_{n+2}\big(\ldots,\Gamma_{A}(p),\bar\Gamma^{A}(q)\big)
  \,&=\,
  {1\over\epsilon} {1\over s_{pq}}
  \left({[p|k_n\!\ket{q} \over 2k_n\cdot(p+q)} - {[p|k_1\!\ket{q} \over 2k_1\cdot(p+q)}\right)
  {\cal M}_n
  + {\cal O}(\epsilon^0),
  \label{Double-soft-N=4SYM-gluino-singlet}
  \\
  {\cal M}_{n+2}\big(\ldots,\phi_{AB}(p),\phi^{AB}(q)\big)
  \,&=\,
  {1\over\epsilon^2} {1\over s_{pq}}
  \left({k_n\cdot(p-q) \over 2k_n\cdot(p+q)} - {k_1\cdot(p-q) \over 2k_1\cdot(p+q)}\right)
  {\cal M}_n
  + {\cal O}(\epsilon^{-1}),
  \label{Double-soft-N=4SYM-scalar-singlet}
  \\
  {\cal M}_{n+2}\big(\ldots,\phi_{AD}(p),\phi^{BD}(q)\big)
  \,&=\,
  {1\over\epsilon}\left({(R_n)^B_{~A} \over 2k_n\cdot(p+q)} - {(R_1)^B_{~A} \over 2k_1\cdot(p+q)}\right)
  {\cal M}_n
  + {\cal O}(\epsilon^0)\,.
  \label{Double-soft-N=4SYM-scalar-nonsinglet}
\end{align}
Similarly for SUGRA, we find that for the following cases of double-soft particles in the supermultiplet \eqref{N=8SG-supermultiplet},
the degenerate solution dominates and we have the corresponding double-soft theorems
\begin{align}
  {\cal M}_{n+2}\big(\ldots,v_{AB}(p),\bar{v}^{AB}(q)\big)
  \,&=\, {\epsilon\over {p\cdot q}}
  \sum_{a=1}^n
  {[p|a\!\ket{q}^2 \over 2k_a\cdot(p+q)}\,
  {\cal M}_n
  + {\cal O}(\epsilon^2),
  \label{Double-soft-N=8SG-graviphoton}
  \\
  {\cal M}_{n+2}\big(\ldots,\chi_{ABC}(p),\bar\chi^{ABC}(q)\big)
  \,&=\, -{1\over s_{pq}}
  \sum_{a=1}^n
  {k_a\cdot(p-q)\, [p|a\!\ket{q} \over 2k_a\cdot(p+q)}\,
  {\cal M}_n
  + {\cal O}(\epsilon),
  \label{Double-soft-N=8SG-graviphotino-singlet}
  \\
  {\cal M}_{n+2}\big(\ldots,\chi_{ADE}(p),\bar\chi^{BDE}(q)\big)
  \,&=\,
  \epsilon \sum_{a=1}^n
  {[p|a\!\ket{q} \over 2k_a\cdot(p+q)}(R_a)^B_{~A}\,
  {\cal M}_n
  + {\cal O}(\epsilon^2),
  \label{Double-soft-N=8SG-graviphotino-nonsinglet}
  \\
  {\cal M}_{n+2}\big(\ldots,\phi_{ABCD}(p),\phi^{ABCD}(q)\big)
  \,&=\, {1\over\epsilon}\, {1 \over s_{pq}}
  \sum_{a=1}^n
  {\big(k_a\cdot(p-q)\big)^2 \over 2k_a\cdot(p+q)}\,
  {\cal M}_n
  + {\cal O}(\epsilon^0),
  \label{Double-soft-N=8SG-scalar-singlet}
  \\
  {\cal M}_{n+2}\big(\ldots,\phi_{ADEF}(p),\phi^{BDEF}(q)\big)
  \,&=\, -
  \sum_{a=1}^n
  {k_a\cdot(p-q) \over 2k_a\cdot(p+q)}
  (R_a)^B_{~A}\, {\cal M}_n
  + {\cal O}(\epsilon).
  \label{Double-soft-N=8SG-scalar-nonsinglet}
\end{align}
Thus we have obtained, from formulas with the 4d rational scattering equations~\cite{Mason-Skinner-1311.2564}, all these universal double-soft theorems, among which some are new and others are known previously. The most famous one is the double soft-scalar theorem \eqref{Double-soft-N=8SG-scalar-nonsinglet} in SUGRA~\cite{Arkani-Hamed-Cachazo-Kaplan-0808.1446}, and more recently, double soft graviphotino (spin-1/2) theorems in supergravity 
were studied in four dimensions as well as three dimensions in \cite{Chen-Huang-Wen-1412.1809,Chen-Huang-Wen-1412.1811}.
In ${\cal N}\!=\!4$ SYM, double scalar theorems~\eqref{Double-soft-N=4SYM-scalar-nonsinglet} were obtained using BCFW recursions in \cite{Volovich-Wen-Zlotnikov-1504.05559,Klose-McLoughlin-Nandan-Plefka-Travaglini-1504.05558}, and from string theory in \cite{Vecchia-Marotta-Mojaza-1507.00938}; double gluino/scalar theorems, \eqref{Double-soft-N=4SYM-gluino-singlet}~and~\eqref{Double-soft-N=4SYM-scalar-singlet}, were given in \cite{Georgiou-1505.08130} from MHV diagrams.

It is also interesting to compare the double soft theorems in different theories.
First let us discuss the case of two soft particles form a SU(${\cal N}$) singlet in supersymmetric theories or without flavors in non-supersymmetric theories.
The double soft scalar factors are all of the form $(p\cdot q)^\alpha\, (k_a\cdot(p-q))^\beta/k_a\cdot(p+q)$, where the exponents $(\alpha,\beta)$ are (0,\,2), (0,\,1), (1,\,2), (-1,\,1), (-1,\,2) for SDBI, NLSM, sGal, SYM and SUGRA respectively.
Similarly double soft factors for spin-$\tfrac12$ fermions are of the form $(p\cdot q)^\alpha\, (k_a\cdot(p-q))^\beta\, [p|k_a\!\ket{q}^\gamma / k_a\cdot(p+q)$, three exponents are (0,\,1,\,1), (-1,\, 0,\, 1) and (-1,\,1,\,1) for SDBI, SYM and SUGRA.
For double soft (gravi-)photon emission, the structure is the same with exponents (0,\,0,\,2) and (-1,\,0,\,2) for SDBI and SUGRA.
For the case that only one flavor index is different in two soft particles, all soft operators involve the R-symmetry SU(${\cal N}$) generator $R_A^B$, and the remaining part has similar structure just like the first case. The similarities of these soft factors may reflect double-copy relations and other connections between the corresponding theories.
We leave it for future study.

\begin{table}[h]
\centering
\begin{tabular}{c|c|c|c|c|c}
\hline
\hline
Theory & Soft particle pair
& ~${\cal O}({\rm d})$~ & ${\cal O}({\rm nd})$ & ${\cal O}({\rm nd})-{\cal O}({\rm d})$
& ``d'' dominant\\ [0.5ex]
\hline
\multirow{5}{*}{${\cal N}\!=\!4$~SDBI}
 & \big($\gamma^+,\, \gamma^-$\big)    & 3 & 4 & 1 & \checkmark \\
 & \big($\psi_A,\, \bar\psi^A$\big)    & 2 & 4 & 2 & \checkmark \\
 & \big($\psi_A,\, \bar\psi^B$\big)    & 3 & 4 & 1 & \checkmark \\
 & \big($\phi_{AB},\, \phi^{AB}$\big)  & 1 & 4 & 3 & \checkmark \\ [0.1ex]
 & \big($\phi_{AD},\, \phi^{BD}$\big)  & 2 & 4 & 2 & \checkmark \\ [0.2ex]
\hline
NLSM & ($\phi,\, \phi$) & 0 & 2  & 2  & \checkmark \\ [0.5ex]
\hline
sGal & ($\phi,\, \phi$) & 3 & 6 & 3 & \checkmark \\ [0.5ex]
\hline
\multirow{5}{*}{${\cal N}\!=\!4$~SYM}
 & \big($g^+,\, g^-$\big)              & 0 & 0 & 0 \\
 & \big($\psi_A,\, \bar\psi^A$\big)    &-1 & 0 & 1 & \checkmark \\
 & \big($\psi_A,\, \bar\psi^B$\big)    & 0 & 0 & 0 &  \\
 & \big($\phi_{AB},\, \phi^{AB}$\big)  &-2 & 0 & 2 & \checkmark \\
 & \big($\phi_{AD},\, \phi^{BD}$\big)  &-1 & 0 & 1 & \checkmark \\ [0.2ex]
\hline
\multirow{9}{*}{${\cal N}\!=\!8$~SUGRA}
 & \big($h^{+},\, h^{-}$\big)               & 3 & 2 & -1 \\
 & \big($\psi_A,\, \bar\psi^A$\big)         & 2 & 2 & 0 \\
 & \big($\psi_A,\, \bar\psi^B$\big)         & 3 & 2 & -1 \\
 & \big($v_{AB},\, \bar{v}^{AB}$\big)       & 1 & 2 & 1 & \checkmark \\
 & \big($v_{AD},\, \bar{v}^{BD}$\big)       & 2 & 2 & 0 \\
 & \big($\chi_{ABC},\, \bar\chi^{ABC}$\big) & 0 & 2 & 2 & \checkmark \\
 & \big($\chi_{ADE},\, \bar\chi^{BDE}$\big) & 1 & 2 & 1 & \checkmark \\
 & \big($\phi_{ABCD},\, \phi^{ABCD}$\big)   &-1 & 2 & 3 & \checkmark \\
 & \big($\phi_{ADEF},\, \phi^{BDEF}$\big)   & 0 & 2 & 2 & \checkmark \\ [0.2ex]
\hline
\hline
\end{tabular}
\caption{Leading scaling in $\epsilon$ of the formulas of scattering amplitudes in the double limit \eqref{soft-limit}.
In soft pairs with flavors indices, one demands $A\ne B$ which corresponds to two soft particles do not form a SU$({\cal N})$ flavor singlet.
Here the tick $\checkmark$ denotes that the degenerate solution is dominant at leading order, and in these cases we give the double-soft theorems in this section. }
\label{Soft-Limits-Amp-table}
\end{table}

\section{Discussions}

In this paper we have studied formulas, inspired by Witten's twistor string \cite{Witten-0312171} and other twistor-string models~\cite{Cachazo-Geyer-1206.6511, Cachazo-Skinner-1207.0741, Cachazo-Mason-Skinner-1207.4712, Skinner-1301.0868, Geyer-Lipstein-Mason-1404.6219}, for four-dimensional tree-level scattering amplitudes in various theories. The formulas are based on 4d scattering equations in either polynomial \eqref{4dp} or rational form \eqref{4dr}, which can be obtained by reducing the general scattering equations \eqref{scatt} to four dimensions. We have shown that the rational-form equations simply follow from fixing the GL$(k)$ redundancy of the polynomial form, and how these two types of formulas for amplitudes are related to each other (see \eqref{integrands}).

What is special and advantageous about working in four dimensions is that the equations and formulas naturally split into sectors. This is not surprising for theories with helicity sectors, such as Yang-Mills, gravity and Born-Infeld theory (with only the helicity-preserving, middle sector). With four-dimensional on-shell superspace, the formulas are most naturally written in supersymmetric form, and in particular we obtain a new formula~\eqref{SDBI} for amplitudes in the ${\cal N}\!=\!4$ supersymmetric completion of DBI~\cite{Bergshoeff-Coomans-Kallosh-Shahbazi-Proeyen-1303.5662}. It is intriguing that formulas for scalars in non-linear sigma model and special Galileon theory only exist in the middle sector, and take a very similar form as that of DBI. This again shows that these scalar theories are very special and have simple amplitudes: the formulas, \eqref{NLSM} and \eqref{sGal}, are in sharp contrast with \eqref{phi3} of $\phi^3$ theory, which requires a sum over all sectors in four dimensions. It is also worth noticing that, the integrands of these formulas \eqref{SDBI}, \eqref{NLSM} and \eqref{sGal}, can be used with both rational and polynomial form of the scattering equations, with the factor $V^{4-\cal N}_{n \over 2}$ for the latter. This is the same as the case of ${\cal N}\!=\!4$ SYM, but not so for ${\cal N}\!=\!8$ SUGRA, bi-adjoint $\phi^3$ etc.

We have applied the formulas to study soft emissions, especially double-soft theorems of amplitudes in these theories. The key idea is the same as in general dimensions~\cite{Cachazo-He-Yuan-1503.04816}, namely universal behavior of double-soft emission is completely controlled by the degenerate solution, see table~\ref{Soft-Limits-Amp-table}. It is remarkable to see that evaluating this solution alone gives all the universal double-soft factors, which in turn provide crucial information on the coset structure of the spontaneous symmetry breaking. In particular, we obtain sub-leading theorems for double-scalar or double-fermion emissions in super-DBI theory, which resemble the double-scalar case in ${\cal N}\!=\!8$ supergravity~\cite{Arkani-Hamed-Cachazo-Kaplan-0808.1446}.
We also classified these double-soft theorems in ${\cal N}\!=\!4$ SYM and ${\cal N}\!=\!8$ SUGRA. From the table, we see that in many cases the degenerate solution is dominant beyond leading order, ${\cal O}({\rm nd})-{\cal O}({\rm d})>1$, such as double-soft scalar emission in ${\cal N}\!=\! 4$ SDBI. In these cases one can derive sub-leading (and even sub-sub-leading) double-soft theorems by the same method as the one in this paper and \cite{Cachazo-He-Yuan-1503.04816}.

A longstanding open question is how to generalize tree-level formulas for ${\cal N}\!=\!4$ SYM and ${\cal N}\!=\!8$ SUGRA, to formulas at one loop. There has been considerable progress for one-loop CHY formulas in general dimensions~\cite{Adamo-Casali-Skinner-1312.3828, Geyer-Mason-Monteiro-Tourkine-1507.00321, Geyer-Mason-Monteiro-Tourkine-1511.06315, Cachazo:2015aol}}, and it would be very interesting to do so for supersymmetric theories in 4d (see \cite{Lipstein:2015vxa} for a conjecture for ${\cal N}\!=\!8$ SUGRA). Another important question is to see what is special about these effective field theories in four dimensions. The supersymmetric DBI theory seems to be a perfect candidate for studying both loop generalizations and the simplicity in 4d. Other interesting directions include further study of the soft theorems and the physics behind it. Just as double-scalar theorems in ${\cal N}\!=\!8$ SUGRA probing the coset structure of $E_{7(7)}$ symmetries, the double-fermion theorems in super-DBI can reveal the structures of non-linearly realized supersymmetries of the theory. Related to this, it would be also very interesting to study sub-leading theorems similar to those in \cite{Cachazo-He-Yuan-1503.04816}, which involve bosonic derivatives (rather than fermionic ones in this paper). Perhaps by combining these two types of sub-leading theorems, one can associate them to possible hidden symmetries and structures.

\section*{Acknowledgments}
S.H.~thanks F.~Cachazo for discussions and hospitality during his visit to Perimeter Institute.  Z.L.~is extremely grateful to Peng Zhang for his generous support during a visit since July 2015 at Renmin University of China, where most of the work was done, and also to Zhengwu Liu, Xiaofeng Pu, Da-Ping Liu and Ke Wang for financial help. J.W.~would like to thank the participants of the advanced workshop ``Dark Energy and Fundamental Theory" supported by the Special Fund for Theoretical Physics from the NSFC with Grant No.~11447613 for stimulating discussion. The work of J.W.~was in part supported by NSFC Grants No.~11222549 and No.~11575202. J.W.~also gratefully acknowledges the support of K.~C.~Wong Education Foundation.


\begin{thebibliography}{100}
\small

\bibitem{Cachazo-He-Yuan-1307.2199}
F. Cachazo, S. He and E. Y. Yuan,
{\it Scattering of Massless Particles in Arbitrary Dimension},
{\it Phys. Rev. Lett.} {\bf 113} (2014) 17, 171601
[\href{http://arxiv.org/abs/1307.2199}{\tt arXiv:1307.2199}]


\bibitem{Cachazo-He-Yuan-1309.0885}
F. Cachazo, S. He and E. Y. Yuan,
{\it Scattering of Massless Particles:~Scalars, Gluons and Gravitons},
{\it JHEP} {\bf 1407} (2014) 033
[\href{http://arxiv.org/abs/1309.0885}{\tt arXiv:1309.0885}]


\bibitem{Cachazo-He-Yuan-1409.8256}
F. Cachazo, S. He and E. Y. Yuan,
{\it Einstein-Yang-Mills Scattering Amplitudes From Scattering Equations},
{\it JHEP} {\bf 1501} (2015) 121
[\href{http://arxiv.org/abs/1409.8256}{\tt arXiv:1409.8256}]


\bibitem{Cachazo-He-Yuan-1412.3479}
F. Cachazo, S. He and E. Y. Yuan,
{\it Scattering Equations and Matrices: From Einstein To Yang-Mills, DBI and NLSM},
{\it JHEP} {\bf 1507} (2015) 149
[\href{http://arxiv.org/abs/1412.3479}{\tt arXiv:1412.3479}]


\bibitem{Hinterbichler-Joyce-1501.07600}
K. Hinterbichler and A. Joyce,
{\it Hidden symmetry of the Galileon},
{\it Phys.~Rev.}~{\bf D 92} (2015) 023503
[\href{http://arxiv.org/abs/1501.07600}{\tt arXiv:1501.07600}]


\bibitem{Hinterbichler:2011tt}
K.~Hinterbichler,
{\it Theoretical Aspects of Massive Gravity},
\href{http://dx.doi.org/10.1103/RevModPhys.84.671}
{{\it Rev.\ Mod.\ Phys.}\  {\bf 84} (2012) 671}
[\href{http://arxiv.org/abs/1105.3735}
{\tt arXiv:1105.3735}]


\bibitem{Dvali:2000hr}
G.~R.~Dvali, G.~Gabadadze and M.~Porrati,
{\it 4-D gravity on a brane in 5-D Minkowski space},
{{\it Phys.\ Lett.}\ {\bf B 485} (2000) 208}
[\href{http://arxiv.org/abs/hep-th/0005016}
{\tt hep-th/0005016}]


\bibitem{Kampf:2014rka}
K.~Kampf and J.~Novotny,
{\it Unification of Galileon Dualities},
{{\it JHEP} {\bf 1410} (2014) 006}
[\href{http://arxiv.org/abs/1403.6813}
{\tt arXiv:1403.6813}]


\bibitem{Cachazo-He-Yuan-1306.2962}
F. Cachazo, S. He and E. Y. Yuan,
{\it Scattering in Three Dimensions from Rational Maps},
{\it JHEP} {\bf 1310} (2013) 141
[\href{http://arxiv.org/abs/1306.2962}{\tt arXiv:1306.2962}]


\bibitem{Cachazo-He-Yuan-1306.6575}
F. Cachazo, S. He and E. Y. Yuan,
{\it Scattering Equations and KLT Orthogonality},
{\it Phys. Rev.} {\bf D 90} (2014) 065001
[\href{http://arxiv.org/abs/1306.6575}{\tt arXiv:1306.6575}]


\bibitem{Fairlie:1972}
D. Fairlie and D. Roberts,
{\it Dual Models without Tachyons\,--\,a New Approach},
unpublished Durham preprint PRINT-72-2440, 1972


\bibitem{*Roberts:1972}
D. Roberts,
{\it Mathematical Structure of Dual Amplitudes},
PhD~thesis, Durham University, 1972
[\href{http://etheses.dur.ac.uk/8662/1/8662_5593.PDF}{\tt available at Durham e-Theses}]


\bibitem{Fairlie-0805.2263}
D. B. Fairlie,
{\it A Coding of Real Null Four-Momenta into World-Sheet Co-ordinates},
{\it Adv. Math. Phys.} {\bf 2009} (2009) 284689
[\href{http://arxiv.org/abs/0805.2263}{\tt arXiv:0805.2263}]


\bibitem{Gross:1987ar}
D. J. Gross and P. F. Mende,
{\it String Theory Beyond the Planck Scale},
{\it Nucl. Phys.} {\bf B 303} (1988) 407


\bibitem{Witten-0403199}
E. Witten,
{\it Parity invariance for strings in twistor space},
{\it Adv. Theor. Math. Phys.} {\bf 8} (2004) 779
[\href{http://arxiv.org/abs/hep-th/0403199}{\tt hep-th/0403199}]


\bibitem{Makeenko-Olesen-1111.5606}
Y. Makeenko and P. Olesen,
{\it The QCD scattering amplitude from area behaved Wilson loops},
{\it Phys. Lett.} {\bf B 709} (2012) 285
[\href{http://arxiv.org/abs/1111.5606}{\tt arXiv:1111.5606}]


\bibitem{Cachazo-1206.5970}
F. Cachazo,
{\it Fundamental BCJ Relation in $N=4$ SYM From The Connected Formulation},
\href{http://arxiv.org/abs/1206.5970}{\tt arXiv:1206.5970}


\bibitem{Berkovits:2013xba}
N.~Berkovits,
{\it Infinite Tension Limit of the Pure Spinor Superstring},
{\it JHEP} {\bf 1403}, 017 (2014)
[\href{http://arxiv.org/abs/1311.4156}{\tt arXiv:1311.4156}]


\bibitem{Mason-Skinner-1311.2564}
L. Mason and D. Skinner,
{\it Ambitwistor strings and the scattering equations},
{\it JHEP} {\bf 1407} (2014) 048
[\href{http://arxiv.org/abs/1311.2564}{\tt arXiv:1311.2564}]


\bibitem{Ohmori-1504.02675}
K. Ohmori,
{\it Worldsheet Geometries of Ambitwistor String},
{\it JHEP} {\bf 1506} (2015) 075
[\href{http://arxiv.org/abs/1504.02675}{\tt arXiv:1504.02675}]


\bibitem{Casali:2015vta}
E.~Casali, Y.~Geyer, L.~Mason, R.~Monteiro and K.~A.~Roehrig,
{\it New Ambitwistor String Theories},
{\it JHEP} {\bf 1511}, 038 (2015)
[\href{http://arxiv.org/abs/1506.08771}{\tt arXiv:1506.08771}].


\bibitem{Roiban-Spradlin-Volovich-0403190}
R. Roiban, M. Spradlin and A. Volovich,
{\it On the tree level S matrix of Yang-Mills theory},
{\it Phys. Rev.} {\bf D 70} (2004) 026009
[\href{http://arxiv.org/abs/hep-th/0403190}{\tt hep-th/0403190}]


\bibitem{Witten-0312171}
E. Witten,
{\it Perturbative gauge theory as a string theory in twistor space},
{\it Commun. Math. Phys.} {\bf 252} (2004) 189
[\href{http://arxiv.org/abs/hep-th/0312171}{\tt hep-th/0312171}]


\bibitem{Cachazo-Geyer-1206.6511}
F. Cachazo and Y. Geyer,
{\it A `Twistor String' Inspired Formula For Tree-Level Scattering Amplitudes in $N=8$ SUGRA},
\href{http://arxiv.org/abs/1206.6511}{\tt arXiv:1206.6511}


\bibitem{Cachazo-Skinner-1207.0741}
F. Cachazo and D. Skinner,
{\it Gravity from Rational Curves in Twistor Space},
{\it Phys. Rev. Lett.} {\bf 110} (2013) 161301
[\href{http://arxiv.org/abs/1207.0741}{\tt arXiv:1207.0741}]


\bibitem{Cachazo-Mason-Skinner-1207.4712}
F. Cachazo, L. Mason and D. Skinner,
{\it Gravity in Twistor Space and its Grassmannian Formulation},
SIGMA {\bf 10} (2014) 051
[\href{http://arxiv.org/abs/1207.4712}{\tt arXiv:1207.4712}]


\bibitem{Skinner-1301.0868}
D. Skinner,
{\it Twistor Strings for $N=8$ Supergravity},
\href{http://arxiv.org/abs/1301.0868}{\tt arXiv:1301.0868}


\bibitem{Geyer-Lipstein-Mason-1404.6219}
Y. Geyer, A. E. Lipstein and L. J. Mason,
{\it Ambitwistor Strings in Four Dimensions},
{\it Phys. Rev. Lett.} {\bf 113} (2014) 081602
[\href{http://arxiv.org/abs/1404.6219}{\tt arXiv:1404.6219}]


\bibitem{Cachazo-Zhang-1601.06305}
F. Cachazo and G. Zhang,
{\it Minimal Basis in Four Dimensions and Scalar Blocks},
\href{http://arxiv.org/abs/1601.06305}{\tt arXiv:1601.06305}


\bibitem{Adamo-Casali-Roehrig-Skinner-1507.02207}
T. Adamo, E. Casali, K. A. Roehrig and D. Skinner,
{\it On tree amplitudes of supersymmetric Einstein-Yang-Mills theory},
{\it JHEP} {\bf 1512} (2015) 177
[\href{http://arxiv.org/abs/1507.02207}{\tt arXiv:1507.02207}]


\bibitem{Tseytlin-9908105}
A. A. Tseytlin,
{\it Born-Infeld action, supersymmetry and string theory},
{\it The many faces of the superworld} (M. A. Shifman ed.) 417
[\href{http://arxiv.org/abs/hep-th/9908105}{\tt hep-th/9908105}]


\bibitem{Bergshoeff-Coomans-Kallosh-Shahbazi-Proeyen-1303.5662}
E. Bergshoeff, F. Coomans, R. Kallosh, C. S. Shahbazi and A. Van Proeyen,
{\it Dirac-Born-Infeld-Volkov-Akulov and Deformation of Supersymmetry},
{\it JHEP} {\bf 1308} (2013) 100
[\href{http://arxiv.org/abs/1303.5662}{\tt 1303.5662}]


\bibitem{Volkov-Akulov-1972}
D. V. Volkov and V. P. Akulov,
{\it Possible universal neutrino interaction},
{\it JETP Lett.} {\bf 16} (1972) 438; {\it Pisma Zh. Eksp. Teor. Fiz.} {\bf 16} (1972) 621


\bibitem{Cachazo-Strominger-1404.4091}
F. Cachazo and A. Strominger,
{\it Evidence for a New Soft Graviton Theorem},
\href{http://arxiv.org/abs/1404.4091}{\tt arXiv:1404.4091}


\bibitem{Adler-1965}
S. L. Adler,
{\it Consistency conditions on the strong interactions implied by a partially conserved axial vector current},
{\it Phys. Rev.} {\bf 137} (1965) B1022


\bibitem{Weinberg-1966-Pino-Scattering}
S. Weinberg,
{\it Pion scattering lengths},
{\it Phys. Rev. Lett.} {\bf 17} (1966) 616


\bibitem{Arkani-Hamed-Cachazo-Kaplan-0808.1446}
N. Arkani-Hamed, F. Cachazo and J. Kaplan,
{\it What is the Simplest Quantum Field Theory?},
{\it JHEP} {\bf 1009} (2010) 016
[\href{http://arxiv.org/abs/0808.1446}{\tt arXiv:0808.1446}]


\bibitem{Cachazo-He-Yuan-1503.04816}
F. Cachazo, S. He and E. Y. Yuan,
{\it New Double Soft Emission Theorems},
{\it Phys. Rev.} {\bf D 92} (2015) 065030
[\href{http://arxiv.org/abs/1503.04816}{\tt arXiv:1503.04816}]


\bibitem{Chen-Huang-Wen-1412.1809}
W.-M. Chen, Y.-t. Huang and C. Wen,
{\it New Fermionic Soft Theorems for Supergravity Amplitudes},
{\it Phys. Rev. Lett.} {\bf 115} (2015) 021603
[\href{http://arxiv.org/abs/1412.1809}{\tt arXiv:1412.1809}]


\bibitem{Britto-Cachazo-Feng-0412308}
R. Britto, F. Cachazo and B. Feng,
{\it New recursion relations for tree amplitudes of gluons},
{\it Nucl. Phys.} {\bf B 715} 499
[\href{http://arxiv.org/abs/hep-th/0412308}{\tt hep-th/0412308}]


\bibitem{Britto-Cachazo-Feng-Witten-0501052}
R. Britto, F. Cachazo, B. Feng and E. Witten,
{\it Direct proof of tree-level recursion relation in Yang-Mills theory},
{\it Phys. Rev. Lett.} {\bf 94} (2005) 181602
[\href{http://arxiv.org/abs/hep-th/0501052}{\tt hep-th/0501052}]


\bibitem{Cachazo-1301.3970}
F. Cachazo,
{\it Resultants and Gravity Amplitudes},
\href{http://arxiv.org/abs/1301.3970}{\tt arXiv:1301.3970}


\bibitem{Arkani-Hamed-Bourjaily-Cachazo-Trnka-0912.4912}
N. Arkani-Hamed, J. Bourjaily, F. Cachazo and J. Trnka,
{\it Unification of Residues and Grassmannian Dualities},
{\it JHEP} {\bf 1101} (2011) 049
[\href{http://arxiv.org/abs/0912.4912}{\tt arXiv:0912.4912}]


\bibitem{Arkani-Hamed-Bourjaily-Cachazo-Trnka-0903.2110}
N. Arkani-Hamed, F. Cachazo, C. Cheung and J. Kaplan,
{\it The S-Matrix in Twistor Space},
{\it JHEP} {\bf 1003} (2010) 110
[\href{http://arxiv.org/abs/0903.2110}{\tt arXiv:0903.2110}]


\bibitem{He-1207.4064}
S. He,
{\it A Link Representation for Gravity Amplitudes},
{\it JHEP} {\bf 1310} (2013) 139
[\href{http://arxiv.org/abs/1207.4064}{\tt arXiv:1207.4064}]


\bibitem{Nair-1988}
V. P. Nair,
{\it A Current Algebra for Some Gauge Theory Amplitudes},
{\it Phys. Lett.} {\bf B 214} (1988) 215


\bibitem{Kawai-Lewellen-Tye-1985}
H. Kawai, D. C. Lewellen and S. H. H. Tye,
{\it A Relation Between Tree Amplitudes of Closed and Open Strings},
{\it Nucl. Phys.} {\bf B269} (1986) 1


\bibitem{Boels-Larsen-Obers-Vonk-0808.2598}
R. Boels, K. J. Larsen, N. A. Obers and M. Vonk,
{\it MHV, CSW and BCFW:~Field theory structures in string theory amplitudes},
{\it JHEP} {\bf 0811} (2008) 015
[\href{http://arxiv.org/abs/0808.2598}{\tt arXiv:0808.2598}]


\bibitem{Luo:2015tat}
H.~Luo and C.~Wen,
{\it Recursion relations from soft theorems},
{\it JHEP} {\bf 1603} (2016) 088
[\href{http://arxiv.org/abs/1512.06801}
{\tt arXiv:1512.06801}]


\bibitem{Cachazo:2016njl}
F.~Cachazo, P.~Cha and S.~Mizera,
{\it Extensions of Theories from Soft Limits},
\href{http://arxiv.org/abs/1604.03893}
{\tt arXiv:1604.03893}


\bibitem{Chen-Huang-Wen-1505.07093}
W.-M. Chen, Y.-t. Huang and C. Wen,
{\it Exact coefficients for higher dimensional operators with sixteen supersymmetries},
{\it JHEP} {\bf 1509} (2015) 098
[\href{http://arxiv.org/abs/1505.07093}{\tt arXiv:1505.07093}]


\bibitem{Klose-McLoughlin-Nandan-Plefka-Travaglini-1504.05558}
T. Klose, T. McLoughlin, D. Nandan, J. Plefka and G. Travaglini,
{\it Double-Soft Limits of Gluons and Gravitons},
{\it JHEP} {\bf 1507} (2015) 135
[\href{http://arxiv.org/abs/1504.05558}{\tt arXiv:1504.05558}]


\bibitem{Volovich-Wen-Zlotnikov-1504.05559}
A. Volovich, C. Wen and M. Zlotnikov,
{\it Double Soft Theorems in Gauge and String Theories},
{\it JHEP} {\bf 1507} (2015) 095
[\href{http://arxiv.org/abs/1504.05559}{\tt arXiv:1504.05559}]


\bibitem{Vecchia-Marotta-Mojaza-1507.00938}
P. Di Vecchia, R. Marotta and M. Mojaza,
{\it Double-soft behavior for scalars and gluons from string theory},
{\it JHEP} {\bf 1512} (2015) 150
[\href{http://arxiv.org/abs/1507.00938}{\tt arXiv:1507.00938}]


\bibitem{Kampf-Novotny-Trnka-1304.3048}
K. Kampf, J. Novotny and J. Trnka,
{\it Tree-level Amplitudes in the Nonlinear Sigma Model},
{\it JHEP} {\bf 1305} (2013) 032
[\href{http://arxiv.org/abs/1304.3048}{\tt arXiv:1304.3048}]


\bibitem{Du-Luo-1505.04411}
Y.-J. Du and H. Luo,
{\it On single and double soft behaviors in NLSM},
{\it JHEP} {\bf 1508} (2015) 058
[\href{http://arxiv.org/abs/1505.04411}{\tt arXiv:1505.04411}]


\bibitem{Low-1512.01232}
I. Low,
{\it Double Soft Theorems and Shift Symmetry in Nonlinear Sigma Models},
{\it Phys. Rev.} {\bf D 93} (2016) 045032
[\href{http://arxiv.org/abs/1512.01232}{\tt arXiv:1512.01232}]


\bibitem{Chen-Huang-Wen-1412.1811}
W.-M. Chen, Y.-t. Huang and C. Wen,
{\it From $U(1)$ to $E_8$:~soft theorems in supergravity amplitudes},
{\it JHEP} {\bf 1503} (2015) 150
[\href{http://arxiv.org/abs/1412.1811}{\tt arXiv:1412.1811}]


\bibitem{Georgiou-1505.08130}
G. Georgiou,
{\it Multi-soft theorems in Gauge Theory from MHV Diagrams},
{\it JHEP} {\bf 1508} (2015) 128
[\href{http://arxiv.org/abs/1505.08130}{\tt arXiv:1505.08130}]


\bibitem{Adamo-Casali-Skinner-1312.3828}
T. Adamo, E. Casali and D. Skinner,
{\it Ambitwistor strings and the scattering equations at one loop},
{\it JHEP} {\bf 1404} (2014) 104
[\href{http://arxiv.org/abs/1312.3828}{\tt arXiv:1312.3828}]


\bibitem{Geyer-Mason-Monteiro-Tourkine-1507.00321}
Y. Geyer, L. Mason, R. Monteiro and P. Tourkine,
{\it Loop Integrands for Scattering Amplitudes from the Riemann Sphere},
{\it Phys. Rev. Lett.} {\bf 115} (2015) 12, 121603
[\href{http://arxiv.org/abs/1507.00321}{\tt arXiv:1507.00321}]


\bibitem{Geyer-Mason-Monteiro-Tourkine-1511.06315}
Y. Geyer, L. Mason, R. Monteiro and P. Tourkine,
{\it One-loop amplitudes on the Riemann sphere},
{\it JHEP} {\bf 1603} (2016) 114
[\href{http://arxiv.org/abs/1511.06315}{\tt arXiv:1511.06315}]


\bibitem{Cachazo:2015aol}
F.~Cachazo, S.~He and E.~Y.~Yuan,
{\it One-Loop Corrections from Higher Dimensional Tree Amplitudes},
\href{http://arxiv.org/abs/1512.05001}{\tt arXiv:1512.05001}


\bibitem{Lipstein:2015vxa}
A.~Lipstein and V.~Schomerus,
{\it Towards a Worldsheet Description of ${\cal N}\!=\!8$ Supergravity},
\href{http://arxiv.org/abs/1507.02936}{\tt arXiv:1507.02936}





\end{thebibliography}
\end{document}